\documentclass[a4paper,11pt]{article}
\pdfoutput=1

\usepackage{jheppub}

\usepackage{amsmath}
\usepackage{amssymb}
\usepackage{bbm}
\usepackage{bm}
\usepackage{color}
\usepackage{graphicx}
\usepackage{hyperref}
\usepackage{mathrsfs}

\let\de=\partial
\DeclareMathOperator{\re}{Re}
\DeclareMathOperator{\tr}{Tr}
\DeclareMathOperator{\sn}{sn}
\DeclareMathOperator{\cs}{cs}
\DeclareMathOperator{\dn}{dn}

\DeclareMathOperator{\Li}{Li}
\newcommand{\imag}{\text{i}}
\newcommand{\Da}{\mathscr{D}}

\newcommand{\Fa}{\mathscr{F}}
\newcommand{\Ha}{\mathscr{H}}
\newcommand{\La}{\mathscr{L}}
\newcommand{\LRG}{\Lambda_\text{RG}}
\newcommand{\dd}{\text{d}}
\newcommand{\he}[1]{#1^\dagger}
\newcommand{\vek}[1]{\bm{#1}}
\newcommand{\ket}[1]{|#1\rangle}
\newcommand{\braket}[2]{\langle#1|#2\rangle}

\title{\boldmath Chiral soliton lattice at next-to-leading order}

\author[a]{Tom\'a\v{s} Brauner}
\author[a,b]{and Helena Kole\v{s}ov\'a}

\affiliation[a]{Department of Mathematics and Physics, University of Stavanger,\\
N-4036 Stavanger, Norway}
\affiliation[b]{Albert Einstein Center for Fundamental Physics, Institute for Theoretical Physics,\\
University of Bern, Sidlerstrasse 5, 3012 Bern, Switzerland}

\emailAdd{tomas.brauner@uis.no}
\emailAdd{kolesova@itp.unibe.ch}

\abstract{We compute the free energy of the chiral soliton lattice state in quantum chromodynamics (QCD) at nonzero baryon chemical potential, temperature and external magnetic field at the next-to-leading order of chiral perturbation theory. This extends previous work where only a special limit of the chiral soliton lattice, the domain wall, was considered. Our results therefore serve as a consistency check of the previously established phase diagram of QCD at moderate magnetic fields and temperature and sub-nuclear baryon chemical potentials. Moreover, we use the result for the free energy to determine the magnetization carried by the domain wall and the chiral soliton lattice, both at the next-to-leading order.}

\dedicated{Dedicated to Ji\v{r}\'{\i} Ho\v{s}ek on the occasion of his 80\textsuperscript{th} birthday.}

\begin{document}

\maketitle
\flushbottom


\section{Introduction}

A combination of a sufficiently strong magnetic field and nonzero baryon chemical potential makes the vacuum of quantum chromodynamics (QCD) unstable with respect to the formation of a periodically modulated neutral pion condensate~\cite{Brauner2017a}. The mechanism behind the formation of this novel state, dubbed chiral soliton lattice (CSL) in analogy with similar phenomena in condensed-matter physics~\cite{Kishine2015a}, relies critically on the chiral anomaly~\cite{Son2008a}, similarly to the notorious two-photon decay of the neutral pion. The prediction of the CSL phase in ref.~\cite{Brauner2017a} was made using the low-energy effective field theory of QCD: the chiral perturbation theory ($\chi$PT). It is therefore model-independent. Its numerical accuracy relies on the leading-order (LO), or tree-level, approximation of $\chi$PT.

The prediction of the CSL phase in the phase diagram of QCD raises a number of immediate questions. The first of these concerns the phenomenological relevance. Nonzero baryon density in combination with zero or low temperature and strong magnetic fields suggests that the CSL state might appear in the cores of neutron stars. Indeed, the average baryon density carried by the CSL lies in the ballpark~\cite{Brauner2017a}. On the other hand, a sufficiently strong magnetic field might be hard to reach by conventional mechanisms. However, the critical magnetic field needed for the formation of CSL might be lowered by a positive-feedback loop arising from interaction with neutrons~\cite{Eto:2012qd}. A similar mechanism for spontaneous magnetization of QCD matter was proposed in ref.~\cite{Yoshiike:2015tha}.

On a more conceptual side, one might also wonder how a solitonic pion state carrying anomaly-induced baryon number might be dynamically created. This is a difficult question, and first works in this direction only appeared recently~\cite{Eto:2022lhu,Higaki:2022gnw}, suggesting nucleation of domain walls from the QCD vacuum similar to bubble nucleation in first-order phase transitions. Another important conceptual question is related to the one-dimensional modulation of the CSL state. Namely, such one-dimensional ordered states should be unstable under thermal fluctuations in the two tranverse directions~\cite{Lee2015a,Hidaka2015b}; this is known as the ``Landau-Peierls instability.'' Strictly speaking, the underlying reason for the instability is the softening of the spectrum of transverse fluctuations, enforced by rotational invariance of the system. The instability can therefore be avoided thanks to the external magnetic field, which breaks rotational invariance explicitly~\cite{Ferrer2020}.

It is nevertheless important to explicitly take into account the effect of thermal fluctuations, not only to rule out the presence of the Landau-Peierls instability, but also to delineate more accurately the domain in the phase diagram of QCD, occupied by the CSL phase. This was done in our previous paper~\cite{Brauner:2021sci}, where we extended the analysis of ref.~\cite{Brauner2017a} to the next-to-leading order (NLO) of the derivative expansion of $\chi$PT. For calculation simplicity, we however only considered the domain wall limit of the general CSL state. This is sufficient to locate the phase transition between the normal and CSL phases, assuming that it occurs via the formation of the domain wall.

The primary goal of this paper is to justify this assumption, and thus to put our previous work~\cite{Brauner:2021sci} on a solid basis. At the same time, we further extend our previous results by computing the magnetization, carried by the CSL. This observable is important for the dynamics of spontaneous generation of strong magnetic fields, and thus completes the list of properties relevant for assessing whether the CSL state might occur in the cores of neutron stars.

The plan of our paper is as follows. In section~\ref{sec:CSLatLO}, we review the properties of the CSL solution at LO, which forms a basis for the subsequent analysis. The strategy and main ingredients of the NLO computation are laid out in section~\ref{sec:setup}. The simpler case of the domain wall is discussed in section~\ref{sec:domainwall}. We fill in some details skipped in ref.~\cite{Brauner:2021sci}, and add the calculation of the magnetization. Section~\ref{sec:CSL} extends the analysis to the general case of CSL; this is the core of the present paper. While all of our calculations are done analytically, the importance of the results is best assessed by illustrating them with concrete numerical values. This is done in section~\ref{sec:results}. In section~\ref{sec:summary} we summarize and append some concluding comments. In order to make the main text of the paper readable, many technical details are relegated to four appendices.


\section{Chiral soliton lattice phase at leading order}
\label{sec:CSLatLO}

We work within the two-flavor version of $\chi$PT, whose LO Lagrangian reads
\begin{equation}
\La=\frac{f_\pi^2}4\left[\tr(D_\mu\he\Sigma D^\mu\Sigma)+2m_\pi^2\re\tr(\Sigma-\mathbbm{1})\right]+\La_\text{WZW}.
\label{LagChPT}
\end{equation}
Here $\Sigma$ is a $2\times 2$ unitary unimodular matrix containing the three pion degrees of freedom. Also, $m_\pi$ and $f_\pi$ are the pion mass and decay constant, respectively, and the latter determines the intrinsic cutoff scale of $\chi$PT as $4\pi f_\pi$. The covariant derivative captures the minimal coupling to the electromagnetic gauge potential $A_\mu$, $D_\mu\Sigma=\de_\mu\Sigma-\imag A_\mu[\tfrac{\tau_3}2,\Sigma]$, with $\tau_3$ being the third Pauli matrix. Finally, the Wess-Zumino-Witten (WZW) term $\La_\text{WZW}$ incorporates the effects of the chiral anomaly~\cite{Wess1971a,Witten1983a} and can be written as~\cite{Son2008a}
\begin{equation}
\La_\text{WZW}=\left(\frac{A_\mu}2-A_\mu^\text{B}\right)j_\text{GW}^\mu,
\label{LagWZW}
\end{equation}
where $A_\mu^\text{B}=(\mu,\vek0)$ is the auxiliary gauge field coupling to baryon number and
\begin{equation}
j^\mu_\text{GW}=-\frac1{24\pi^2}\epsilon^{\mu\nu\alpha\beta}\tr\left[ (\Sigma D_\nu\he\Sigma)(\Sigma D_\alpha\he\Sigma)(\Sigma D_\beta\he\Sigma) +\frac{3\imag}4F_{\nu\alpha}\tau_3(\Sigma D_\beta\he\Sigma+D_\beta\he\Sigma\Sigma)\right]
\end{equation}
is the topological Goldstone-Wilczek current~\cite{Goldstone1981a}.

In sufficiently strong magnetic fields, $B\gtrsim m_\pi^2$, charged pions get a large effective mass due to Landau level quantization and only neutral pions remain as relevant degrees of freedom. Defining a dimensionless neutral pion field as $\phi\equiv \pi_0/f_\pi$, $\Sigma=e^{\imag\tau_3\phi}$ can be then inserted in eq.~\eqref{LagChPT}, leading to
\begin{equation}
\La_\phi=\frac{f_\pi^2}2(\de_\mu\phi)^2+m_\pi^2f_\pi^2(\cos\phi-1)+\frac{\mu}{4\pi^2}\vek B\cdot\vek\nabla\phi,
\label{Lagrangian}
\end{equation}
where $\vek B$ denotes the background magnetic field, assumed to be constant and uniform from now on.

The classical ground state of the theory is found by subsequent minimization of the corresponding Hamiltonian with respect to $\phi$. Let us note that the unit matrix $\mathbbm{1}$ was included in the second term of~\eqref{LagChPT} to ensure that the ``normal phase'' (that is, the QCD vacuum, corresponding to $\Sigma_0=\mathbbm{1}$) has zero energy density. A negative (spatial average of) energy density is then a smoking gun for a nontrivial ground state.


\subsection{Classical ground state}

As shown in ref.~\cite{Brauner2017a}, the equation of motion corresponding to the Lagrangian~\eqref{Lagrangian} has a class of topologically nontrivial solutions, modulated in the direction of the magnetic field. We choose to orient $\vek B$ along the $z$-axis. The solutions, $\phi_0$, are parameterized by the dimensionless elliptic modulus $k$ ($0\leq k \leq 1$), and given implicitly by
\begin{equation}
\cos\frac{\phi_0(\bar{z})}2=\sn(\bar{z},k).
\label{solCSL}
\end{equation}
Here $\bar{z}\equiv zm_\pi/k$ is a suitably chosen dimensionless coordinate and $\sn$ is one of the Jacobi elliptic functions. The solution $\phi_0$ is quasiperiodic in $z$ in the sense that
\begin{equation}
\phi_0(z+L)=\phi_0(z)+2\pi,
\end{equation}
where
\begin{equation}\label{CSLperiod}
L=\frac{2k K(k)}{m_\pi}
\end{equation}
is the period and $K(k)$ is the complete elliptic integral of the first kind. As a consequence, the corresponding unitary matrix $\Sigma_0=e^{\imag\tau_3\phi_0}$, which in turn determines the expectation values of the quark scalar and pseudoscalar bilinears (condensates), is strictly periodic.

The average energy density of the CSL solution~\eqref{solCSL} at LO reads
\begin{equation}
\frac{\Fa_\text{0,CSL}}V=2m_\pi^2f_\pi^2\biggl[1-\frac1{k^2}+\frac2{k^2}\frac{E(k)}{K(k)}\biggr]-\frac{m_\pi\mu B}{4\pi kK(k)}.
\label{ECSL}
\end{equation}
The energy density can in turn be minimized with respect to $k$ in order to find the ground state for given baryon chemical potential $\mu$ and magnitude of the external magnetic field $B$. This leads to an implicit condition that uniquely fixes the value of $k$,
\begin{equation}\label{condk}
\frac{E(k)}{k}=\frac{\mu B}{16\pi m_\pi f_\pi^2},
\end{equation}
where $E(k)$ is the complete elliptic integral of the second kind. It follows from eq.~\eqref{ECSL} that whenever the condition~\eqref{condk} is satisfied with $0<k<1$, the LO energy of the CSL state is negative, i.e., CSL is favored over the QCD vacuum. However, eq.~\eqref{condk} cannot be satisfied for arbitrary values of $\mu$ and $B$, since its left-hand side is bounded from below, $E(k)/k\geq 1$. The condition for the existence of the CSL solution can be cast as a lower bound on the magnetic field (for given $\mu$),
\begin{equation}
B \geq B_{\text{CSL}} \equiv \frac{16\pi m_\pi f_\pi^2}{\mu}.
\label{BCSL_LO}
\end{equation}
The equation $B=B_\text{CSL}$ defines the boundary between the normal and CSL phases at LO of the derivative expansion of $\chi$PT.

At the phase transition, the optimum value of the elliptic modulus is $k\to1$. In this limit, the lattice spacing~\eqref{CSLperiod} diverges. The CSL solution~\eqref{solCSL} turns into a domain wall, which has a simple expression not requiring Jacobi elliptic functions,
\begin{equation}
\phi_0(z)=4\arctan e^{m_\pi z}.
\label{solwall}
\end{equation}
The domain wall solution is localized in the direction of the magnetic field, and its energy thus does not scale with volume but rather with area in the two transverse directions. While its average bulk energy density~\eqref{ECSL} vanishes, the domain wall still carries nonzero energy per unit transverse area, which vanishes for $B\to B_\text{CSL}$,
\begin{equation}
\frac{\Fa_\text{0,wall}}S=8m_\pi f_\pi^2-\frac{\mu B}{2\pi}.
\label{Ewall}
\end{equation}


\subsection{Power counting}

In order to include the contributions of quantum and thermal fluctuations to the free energy in a controlled way, a power-counting scheme is needed. We modify the standard $\chi$PT power counting~\cite{Scherer2012a} in the same way as in ref.~\cite{Brauner:2021sci}, i.e., we assign an unconventional counting order to the baryon chemical potential,
\begin{equation}
\partial_\mu,\,m_\pi,\,T,\, A_\mu = \mathcal{O}(p^1),\quad A_\mu^B = \mathcal{O}(p^{-1}). 
\end{equation}
The advantage of this scheme is that it brings the term proportional to $A_\mu^B$ in the WZW Lagrangian~\eqref{LagWZW} to the LO, $\mathcal{O}(p^2)$ part of the effective Lagrangian. This makes the tree-level analysis of CSL in ref.~\cite{Brauner2017a} consistent. 

Higher orders of the derivative expansion of the free energy density of $\chi$PT are then organized by increasing powers of the dimensionless expansion parameter $\epsilon\sim [p/(4\pi f_\pi)]^2$, where $p$ is the characteristic scale of $m_\pi$, $T$ or $\sqrt{B}$. In order to determine the complete free energy at NLO, $\mathcal{O}(p^4)$, we need to include one-loop contributions generated by the LO Lagrangian as well as tree-level contributions from additional, $\mathcal{O}(p^4)$ operators in the effective Lagrangian.

Note that the baryon chemical potential does not correspond to any physical mass scale in $\chi$PT. Consequently, assigning $\mu$ a counting order of $-1$ does not amount to restricting to a particular range of values of $\mu$. However, eq.~\eqref{Lagrangian} shows that the combination $\mu B/(4\pi^2f_\pi^2)$ controls the gradient of the neutral pion field in the CSL state. This is consistent with the fact that $\mu B$ is $\mathcal{O}(p^1)$ according to our counting scheme. It can be checked that the gradient of the neutral pion field in the CSL state given by eqs.~\eqref{solCSL} and~\eqref{condk}  is indeed smaller than 4$\pi f_\pi$ in the relevant part of the parameter space.


\section{Setup of the next-to-leading-order calculation}
\label{sec:setup}

\subsection{Expansion of free energy}\label{sec:expF}

Upon including the effects of fluctuations, the free energy becomes a nonlocal functional of the field configuration $\phi$. Minimizing such a functional directly to find the ground (equilibrium) state would be a daunting task. Fortunately, in order to pin down the phase diagram, we do not really need to know the exact CSL ground state, merely the difference of its free energy compared to the normal phase. Let us see how far we can get by using the systematic power counting of $\chi$PT.

In order to make our task feasible, we will restrict to the class of quasiperiodic functions $\phi(z)$. Each such function can be represented by a profile function $\varphi$ and the period $L$ as
\begin{equation}
\phi(z)=\varphi(z/L).
\end{equation}
The profile function is likewise quasiperiodic, but its period is rescaled to unity. Now recall that already at LO, the procedure for finding the ground state was split into two steps. In the first step, we found a broad class of stationary field configurations of the classical action. In the second step, we then minimized the energy of these solutions with respect to a single real parameter, $k$. With this in mind, we temporarily treat the average free energy density as a functional of the profile $\varphi$ and a function of the period $L$, $\Fa[\varphi,L]$. In the derivative expansion, the free energy of $\chi$PT can be expanded in powers of $\epsilon$,
\begin{equation}
\Fa[\varphi,L]=\Fa_0[\varphi,L]+\epsilon^1\Fa_1[\varphi,L]+\epsilon^2\Fa_2[\varphi,L]+\dotsb.
\label{Fpertexp}
\end{equation}
Denoting the ground state configuration with a bar to distinguish it from a generic test field, it can likewise be expanded in a power series,
\begin{equation}
\begin{split}
\bar\varphi&=\varphi_0+\epsilon^1\varphi_1+\epsilon^2\varphi_2+\dotsb,\\
\bar L&=L_0+\epsilon^1L_1+\epsilon^2L_2+\dotsb.
\end{split}
\label{phiLbar}
\end{equation}
Up to the NLO, that is first order in $\epsilon$, the free energy density then becomes
\begin{equation}
\Fa[\bar\varphi,\bar L]=\Fa_0[\varphi_0,L_0]+\epsilon^1\int\dd z\,\varphi_1(z)\left.\frac{\delta\Fa_0}{\delta\varphi(z)}\right\rvert_{\substack{\varphi=\varphi_0\\ L=L_0}}+\epsilon^1L_1\left.\frac{\de\Fa_0}{\de L}\right\rvert_{\substack{\varphi=\varphi_0\\ L=L_0}}+\epsilon^1\Fa_1[\varphi_0,L_0]+\mathcal O(\epsilon^2).
\label{FNLO}
\end{equation}

In order that the expansion~\eqref{phiLbar} actually represents the ground state, the LO approximation must be a stationary state of $\Fa_0$, that is one of the solutions~\eqref{solCSL}. This ensures vanishing of the integral term in eq.~\eqref{FNLO}; the fact that $\varphi(z)$ has a fixed boundary condition and a fixed period guarantees the vanishing of the boundary terms generated by integration by parts when evaluating the variation $\delta\Fa_0/\delta\varphi$. On the class of solutions~\eqref{solCSL}, eq.~\eqref{FNLO} therefore reduces to
\begin{equation}
\Fa[\bar\varphi,\bar L]=\Fa_0[\varphi_0,L_0]+\epsilon^1L_1\left.\frac{\de\Fa_0}{\de L}\right\rvert_{\substack{\varphi=\varphi_0\\ L=L_0}}+\epsilon^1\Fa_1[\varphi_0,L_0]+\mathcal O(\epsilon^2).
\end{equation}
This enables us to compute the NLO free energy of the LO ground state for given magnetic field and chemical potential, for which also the derivative $\de\Fa_0/\de L$ vanishes. The calculation of the contribution $\Fa_1[\varphi_0,L_0]$ is the main subject of this work.


\subsection{Fluctuation determinant}\label{sec:FluctDet}

In any scalar field theory, the one-loop contribution to the free energy is given by the determinant of the differential operator that defines the part of the Lagrangian quadratic in fluctuations around the chosen ground state. Since the CSL state conserves electric charge, we can consider separately the fluctuations of the neutral and charged pion fields. Eventually, the one-loop free energy will be given by the formal expression
\begin{equation}
\beta\Fa_1^\text{1-loop}=\frac12\tr\log\Da^{(\pi^0)}+\tr\log\Da^{(\pi^\pm)},
\label{defF1loop}
\end{equation}
where $\Da^{(\pi^0)}$ and $\Da^{(\pi^\pm)}$ are the fluctuation operators, specified concretely below. The trace is taken over the spectrum of these operators, hence, detailed information about this spectrum and the density of states will be needed.

In order to find the fluctuation operators $\Da^{(\pi^0)}$ and $\Da^{(\pi^\pm)}$, one needs to choose a convenient parameterization of the matrix variable $\Sigma$, and to expand the Lagrangian~\eqref{LagChPT} to second order in the pion fields. This calculation was performed in appendix~A of ref.~\cite{Brauner2017a} and the fluctuation operator related to neutral pion fluctuations is easy to read off,
\begin{equation}
\Da^{(\pi^0)}=\Box+m_\pi^2\cos\phi_0.
\label{FDet_pi0}
\end{equation}
For the fluctuation operator in the charged pion sector, we need to choose a gauge for the external magnetic field. With the standard asymmetric gauge, $\vek A=(0,Bx,0)$, the fluctuation operator becomes (see appendix~C of ref.~\cite{Brauner2017a} for details)
\begin{equation}
\Da^{(\pi^\pm)}=\Box+2\imag Bx\de_y+B^2x^2-(\phi_0')^2+m_\pi^2\cos\phi_0,
\label{FDet_charged}
\end{equation}
where the prime indicates a derivative with respect to $z$. Note that we are neglecting here the contribution of the WZW term to the bilinear Lagrangian as it is of higher order when the magnetic field is also counted as a small expansion parameter.


\subsubsection{Neutral pions}

Combining equations~\eqref{defF1loop},~\eqref{FDet_pi0} and the CSL solution~\eqref{solCSL}, the contribution of neutral pions to the one-loop free energy of the CSL state follows. Upon analytical continuation to imaginary time and Fourier transformation to the frequency-momentum space, it reads
\begin{equation}
\label{F1CSL}
\beta\Fa^{\text{1-loop},(\pi^0)}_\text{1,CSL}=\frac12\sum_{n,\vek p_\perp,\lambda}\log\left[\omega_n^2+\vek p_\perp^2+m_\pi^2\left(\frac\lambda{k^2}-1\right)\right].
\end{equation}
Here $\omega_n$ stands for the $n$-th bosonic Matsubara frequency, $\omega_n\equiv2\pi nT$, $\vek p_\perp$ is the two-vector of momentum in the directions transverse to the magnetic field, and $\lambda$ labels the (dimensionless) eigenvalues of the effective one-dimensional fluctuation Hamiltonian
\begin{equation}
\Ha_\text{CSL}^{(\pi^0)}\equiv-\de_{\bar{z}}^2+2k^2\sn^2(\bar{z},k). 
\label{HamCSL}
\end{equation}
This is a special case of the so-called Lam\'e Hamiltonian, whose spectrum is known in the literature. We provide some further details with references in appendix~\ref{app:spectra}.

The expressions for the one-loop neutral pion free energy and effective one-dimensional Hamiltonian in the case of the domain wall can in principle be obtained from eqs.~\eqref{F1CSL} and~\eqref{HamCSL} by taking the limit $k\to1$. However, it turns out convenient to use another definition of the eigenvalue $\lambda$, differing from that in eq.~\eqref{F1CSL} by a constant shift,
\begin{equation}
\beta\Fa^{\text{1-loop},(\pi^0)}_\text{1,wall}=\frac12\sum_{n,\vek p_\perp,\lambda}\log[\omega_n^2+\vek p_\perp^2+m_\pi^2(1+\lambda)].
\label{F1wall}
\end{equation}
With this definition, $\lambda$ runs over the eigenvalues of the one-dimensional Hamiltonian
\begin{equation}
\Ha_\text{wall}^{(\pi^0)}\equiv-\de_{\bar{z}}^2-\frac2{\cosh^2\bar{z}}.
\label{Hamwall}
\end{equation}
Here $\bar{z}\equiv m_\pi z$ is the dimensionless coordinate appropriate for the domain wall solution and eq.~\eqref{Hamwall} is a special case of the so-called P\"oschl-Teller Hamiltonian. Its spectrum is also well-known; see appendix~\ref{app:spectra} for a detailed derivation.


\subsubsection{Charged pions}

In the case of charged pions, the transverse directions cannot be dealt with by mere Fourier transformation. Instead, we expect the different eigenstates to organize into Landau levels, labeled by a non-negative integer quantum number $m$. Since the parts of $\Da^{(\pi^\pm)}$ acting on the $x,y$ and $z$ coordinates mutually commute, the Landau level problem can be solved in the usual manner regardless of the background $\phi_0(z)$. The final expression for the charged pion contribution to the one-loop free energy of the CSL state thus parallels eq.~\eqref{F1CSL},
\begin{equation}
\label{F2CSL}
\beta\Fa^{\text{1-loop},(\pi^\pm)}_\text{1,CSL}=\frac{BS}{2\pi}\sum_{n,m,\lambda}\log\left[\omega_n^2+(2m+1)B +\frac{m_\pi^2}{k^2}(\lambda-k^2-4)\right].
\end{equation}
We used the fact that the number of states in a given Landau level per unit transverse area equals $B/2\pi$. The index $\lambda$ now runs over the eigenvalues of the dimensionless Hamiltonian
\begin{equation}
\Ha_\text{CSL}^{(\pi^\pm)}\equiv-\de_{\bar{z}}^2+6k^2\sn^2(\bar{z},k).
\label{HamCSLcharged}
\end{equation}
For the domain wall ($k=1$), we again change the definition of $\lambda$ by shifting it by a constant. This leads to
\begin{equation}
\label{F2wall}
\beta\Fa^{\text{1-loop},(\pi^\pm)}_\text{1,wall}=\frac{BS}{2\pi}\sum_{n,m,\lambda}\log[\omega_n^2+(2m+1)B+m_\pi^2(1+\lambda)],
\end{equation}
with the corresponding one-dimensional effective Hamiltonian
\begin{equation}
\Ha_\text{wall}^{(\pi^\pm)}\equiv-\de_{\bar{z}}^2-\frac6{\cosh^2\bar{z}}.
\label{Hamwallcharged}
\end{equation}
The Hamiltonians~\eqref{HamCSLcharged} and~\eqref{Hamwallcharged} are again special cases of the class of Lam\'e and P\"oschl-Teller Hamiltonians, respectively, and their spectra are described in appendix~\ref{app:spectra}. With all the spectra of the one-dimensional fluctuation Hamiltonians known, the task to evaluate the one-loop free energy of the CSL and domain wall solutions reduces to finding an efficient way to perform the different sums.


\subsection{Renormalization at one loop}
\label{sec:renormalization}

Before proceeding to a detailed calculation, note that the naive one-loop free energy~\eqref{defF1loop} is divergent and requires renormalization. Within the power counting of $\chi$PT, this amounts to adding a new set of NLO operators carrying the necessary counterterms. In the two-flavor version of $\chi$PT, the NLO Lagrangian contains altogether 12 independent operators, see for instance section~3.5.1 of ref.~\cite{Scherer2012a}. Fortunately, we do not need to carry out a complete one-loop renormalization of $\chi$PT. All we need is renormalization of the one-loop effective action on a \emph{neutral} pion background. Moreover, we are only interested in the difference of the free energies of the CSL and normal phases, which means that we can subtract the effective action evaluated on the trivial vacuum, $\phi_0=0$. This reduces the required counterterm Lagrangian to
\begin{equation}
\La_\text{c.t.}= \ell_1[(\de_\mu\phi)^2]^2+\ell_2(\de_\mu\phi)^2m_\pi^2\cos\phi+\ell_3m_\pi^4(\cos^2\phi-1).
\label{Lct}
\end{equation}
where $\ell_{1,2,3}$ are dimensionless couplings that can be fixed independently of the background, that is, with the help of existing $\chi$PT results. In particular, rewriting the couplings $\ell_{1,2,3}$ using the notation of~\cite{Andersen2012b}, we find
\begin{equation}
\ell_1=l_1+l_2,\qquad
\ell_2=l_4,\qquad
\ell_3=l_3+l_4.
\label{Ltoell}
\end{equation}
The divergent parts of $l_{1,2,3,4}$ can in turn be fixed in the vacuum using the $\overline{\text{MS}}$ renormalization scheme with dimensional regularization in $D\equiv4-2\epsilon$ spacetime dimensions. If the renormalization scale is chosen as $m_\pi$,\footnote{There is no residual dependence of our results on the choice of renormalization scale. Of course, one-loop corrections to the free energy generate logarithmic dependence on the renormalization scale, but this can be \emph{exactly} compensated by suitable running of the NLO couplings. This is eventually because the predictions of $\chi$PT are organized by powers of derivatives rather than couplings, and because the NLO free energy is strictly linear in the NLO couplings.} they read~\cite{Andersen2012b} 
\begin{equation}
l_i=-\frac{\gamma_i}{2(4\pi)^2}\left(\frac1\epsilon+1-\bar l_i\right),
\label{MSbar}
\end{equation}
with the algebraic coefficients $\gamma_i$ given by~\cite{Gasser1984a}
\begin{equation}
\gamma_1=\frac13,\qquad
\gamma_2=\frac23,\qquad
\gamma_3=-\frac12,\qquad
\gamma_4=2.
\end{equation}
As for the finite parts of the counterterms, we assume the following values,
\begin{equation}
\bar l_1=-0.4\pm 0.6,\qquad \bar l_2=4.3 \pm 0.1,\qquad \bar l_3=3.53\pm 0.26,\qquad \bar l_4=4.4\pm 0.2.
\label{Lbar}
\end{equation}
The values of $\bar l_{1,2,4}$ were determined by matching of $\chi$PT to experimental input at intermediate energies via Roy equations (see refs.~\cite{Ananthanarayan:2000ht,Colangelo:2001df} for details). On the other hand, the error of $\bar{l}_3$ can be substantially reduced by taking into account lattice simulations. We therefore use the value of $\bar{l}_3$ based on $N_\text{f}=2+1+1$ simulations~\cite{Baron:2011sf,Aoki:2019cca}. For the sake of producing our own numerical results, we will only use the mean values of $\bar l_{1,2,3,4}$; we display the errors in eq.~\eqref{Lbar} just for a rough indication of the uncertainty of our results.

Another important consequence of the one-loop corrections to the effective action of $\chi$PT is that beyond LO, the parameters $f_\pi$ and $m_\pi$ in eq.~\eqref{LagChPT} no longer have the interpretation as the pion decay constant and mass, respectively. In order to be able to match these parameters correctly to physical observables, the part of the full 1-loop-renormalized effective Lagrangian bilinear in the neutral pion field has to be found and compared to the expression $\frac{F_\pi^2}2(\de_\mu\phi)^2-\frac12M_\pi^2F_\pi^2\phi^2$. A detailed calculation then allows for the matching,
\begin{align}\label{fpi}
F_\pi^2&=f_\pi^2+\frac{m_\pi^2}{8\pi^2}\bar l_4,\\
M_\pi^2F_\pi^2&=m_\pi^2f_\pi^2-\frac{m_\pi^4}{32\pi^2}(\bar l_3-4\bar l_4).\label{mpifpi2}
\end{align}
Combining the physical input $M_\pi=140\text{ MeV}$ and $F_\pi=92\text{ MeV}$ with the mean values of the finite counterterms~\eqref{Lbar} gives
\begin{equation}
m_\pi\approx142\text{ MeV},\qquad
f_\pi\approx86\text{ MeV}.
\label{NLOparameters}
\end{equation}
We use these parameter values below in section~\ref{sec:results} to map the part of the phase diagram where the CSL is the ground state.


\section{Domain wall at next-to-leading order}
\label{sec:domainwall}

A complete expression for the NLO free energy of the domain wall appeared in ref.~\cite{Brauner:2021sci}. However, only the basic steps of the calculation were outlined therein, owing to space restrictions of the letter format. Here we review the derivation in full detail. In addition, we derive formulas for domain wall magnetization that were not published previously.


\subsection{One-loop free energy}\label{sec:wall_FNLO}

Let us start with evaluating explicitly the contribution to the free energy from the counterterm Lagrangian~\eqref{Lct}. To that end, we need the following elementary integrals, valid for the domain wall solution~\eqref{solwall},
\begin{equation}
\begin{split}
\int_{-\infty}^{+\infty}\dd z\,[\phi_0'(z)]^4&=\frac{64m_\pi^3}3,\\
\int_{-\infty}^{+\infty}\dd z\,[\phi_0'(z)]^2\cos\phi_0(z)&=-\frac{8m_\pi}3,\\
\int_{-\infty}^{+\infty}\dd z\,[\cos^2\phi_0(z)-1]&=-\frac8{3m_\pi}.
\end{split}
\label{CTwall}
\end{equation}
The counterterm Lagrangian~\eqref{Lct} then gives the following contribution to the free energy of the domain wall per unit transverse area,
\begin{equation}
\frac{\Fa_\text{1,wall}^\text{c.t.}}S=m_\pi^3\left(-\frac{64}3\ell_1-\frac83\ell_2+\frac83\ell_3\right).
\label{Fctwall}
\end{equation}

With the couplings $\ell_{1,2,3}$ fixed according to section~\ref{sec:renormalization}, this contribution serves to cancel the divergences in the one-loop zero-temperature free energy that we shall calculate next. Before doing so, let us outline the general strategy for evaluation of the sums over $\lambda$ in eqs.~\eqref{F1wall} and~\eqref{F2wall}. Using the known properties of the spectrum and eigenstates of the P\"oschl-Teller Hamiltonian, one can derive the summation master formulas (see appendix~\ref{app:spectra} for details)
\begin{equation}
\sideset{}{'}\sum_\lambda f(\lambda)=f(-1)-\frac2\pi\int_0^\infty\frac{f(P^2)}{1+P^2}\,\dd P
\label{lambdasumneutral}
\end{equation}
for the Hamiltonian~\eqref{Hamwall}, governing neutral pion fluctuations, and
\begin{equation}
\label{lambdasumcharged}
\sideset{}{'}\sum_\lambda f(\lambda)={}f(-4)+f(-1)-\frac1\pi\int_0^\infty\left(\frac2{1+P^2}+\frac4{4+P^2}\right)f(P^2)\,\dd P
\end{equation}
for the Hamiltonian~\eqref{Hamwallcharged}, governing charged pion fluctuations. Both formulas hold for any smooth function $f$ satisfying $\lim\limits_{P\to\infty}f(P^2)/P=0$. The primes on the summation symbols indicate regularization of the sums by subtraction of an analogous sum over the spectrum of the ``free particle'' Hamiltonian $\Ha_0 = -\partial^2_{\bar z}$. If these formulas are used in eqs.~\eqref{F1wall} and~\eqref{F2wall}, the difference of free energies of the domain wall and the normal phase is obtained. If not stated otherwise, we will have in mind this difference when speaking about the free energy of the domain wall or CSL in the following text, i.e., the subtraction of the normal phase will not be mentioned explicitly every time.


\subsubsection{Zero-temperature part}\label{sec:Fwall0}

The one-loop free energy can be split into a zero-temperature part and a thermal part in a standard manner. We will first deal with the zero-temperature part, which is somewhat easier to evaluate but requires renormalization. Starting with the contribution of neutral pions, a combination of eqs.~\eqref{F1wall} and~\eqref{lambdasumneutral} gives
\begin{equation}\label{FT0NeutralWall1}
\frac{\Fa_\text{1,wall}^{T=0,(\pi^0)}}S={}\frac12 \left(\frac{e^{\gamma_\text{E}}\Lambda_\text{RG}^2}{4\pi}\right)^\epsilon \int\frac{\dd^dP_\perp}{(2\pi)^d}\biggl\{\log P_\perp^2-\frac1\pi\int_{-\infty}^{+\infty}\dd P\,\frac{\log[P_\perp^2+m_\pi^2(1+P^2)]}{1+P^2}\biggr\}.
\end{equation}
Here $P_\perp$ denotes the transverse components of four-momentum (that is frequency $\omega$ that becomes continuously-valued in the limit $T\to0$, and transverse momentum $\vek p_\perp$), and $d\equiv 3-2\epsilon$ is the number of spatial dimensions in dimensional regularization. As usual in dimensional regularization, we have inserted an appropriate power of the renormalization scale $\Lambda_\text{RG}$ to keep the mass dimension of the free energy fixed. As mentioned in section~\ref{sec:renormalization}, we set $\Lambda_\text{RG} = m_\pi$ in our final results. Recall also that the finite parts of the counterterms are fixed within the $\overline{\text{MS}}$ renormalization scheme. In order to ensure the corresponding subtraction in our calculation, we follow the convention of ref.~\cite{Andersen2012b} and introduce the factor $(e^{\gamma_\text{E}}/4\pi)^\epsilon$, with $\gamma_\text{E}$ being the Euler constant. 

The integral of $\log P_\perp^2$ does not contain any scale and therefore vanishes in dimensional regularization. The remaining contribution is evaluated by first integrating over $P_\perp$ and then over $P$, the final result being
\begin{equation}
\frac{\Fa_\text{1,wall}^{T=0,(\pi^0)}}S=-\frac2d\frac{(e^{\gamma_\text{E}}\Lambda_\text{RG}^2)^{\epsilon}\,m_\pi^d}{(4\pi)^2}\,\Gamma(\tfrac{1-d}2).
\label{FT0wallneutral}
\end{equation}

The contribution of charged pions is considerably more involved due to Landau level quantization. The starting point is eq.~\eqref{F2wall}, where at zero temperature the Matsubara sum has to be replaced with
\begin{equation}\label{repOmega}
T\sum_n\to\int\frac{\dd^{d-2}\omega}{(2\pi)^{d-2}}.
\end{equation}
Note that we treat $\omega$ as living in $d-2$ dimensions for compatibility with dimensional regularization, since the P\"oschl-Teller Hamiltonian~\eqref{Hamwallcharged} is strictly one-dimensional, and the Landau-levels to be summed over describe quantized eigenstates in two transverse directions. With this in mind, we use eq.~\eqref{lambdasumcharged} and subsequently integrate over $\omega$ to find
\begin{align}\notag
\frac{\Fa_\text{1,wall}^{T=0,(\pi^\pm)}}{S} =-&\frac{B(e^{\gamma_\text{E}}\LRG^2)^\epsilon}{2\pi}\,\frac{\Gamma(1-\frac d2)}{\sqrt{4\pi}}\sum_{m=0}^\infty\biggl\{\bigl[(2m+1)B-3m_\pi^2\bigr]^{\frac d2-1}+\bigl[(2m+1)B\bigr]^{\frac d2-1}\\
\label{F0wallchargedaux}
&-\int_0^\infty\frac{\dd P}{2\pi}\left[\frac4{1+P^2}+\frac8{4+P^2}\right]\bigl[(2m+1)B+m_\pi^2(1+P^2)\bigr]^{\frac d2-1}\biggr\}.
\end{align}

The next step is to sum over the Landau levels, which is done with the help of the Hurwitz $\zeta$-function,
\begin{equation}\label{HurwitzDef}
\zeta(s,q)\equiv\sum_{m=0}^\infty\frac1{(m+q)^s}.
\end{equation}
Specifically, we need sums of the form
\begin{equation}
\sum_{m=0}^\infty\bigl[(2m+1)B+x\bigr]^{-s}=(2B)^{-s}\zeta(s,\tfrac12+\tfrac x{2B}).
\label{Hurwitz1}
\end{equation}
In order to be able to extract the divergent part of the free energy analytically, we also need the asymptotic expansion of the sum over Landau levels at large longitudinal momentum $P$. To that end, one can use the so-called Hermite formula for the Hurwitz $\zeta$-function, see section~13.2 of ref.~\cite{Whittaker1927a}, to deduce the following expansion at large $q$,
\begin{equation}
\zeta(s,q+v)=\frac{q^{1-s}}{s-1}+q^{-s}\left(\frac12-v\right)+sq^{-1-s}\left[\frac{1}{12}+\frac{v}{2}(v-1)\right]+\mathcal O(q^{-2-s}).
\label{Hurwitz2}
\end{equation}
We use this formula to expand eq.~\eqref{F0wallchargedaux} summed over $m$ in powers of $1+P^2$ and $4+P^2$. This makes it possible to evaluate the divergent part of the one-loop free energy of charged pions in a closed form. 

When combined with~\eqref{FT0wallneutral}, one can check explicitly that the divergence of the one-loop free energy of the domain wall is canceled by the counterterms~\eqref{Fctwall}. This is an important consistency check of our calculation. The final form of the remaining finite, renormalized NLO zero-temperature free energy of the domain wall is
\begin{align}
\notag
\frac{\Fa_\text{1,wall}^{T=0}}S=&-\frac{m_\pi^3}{72\pi^2}&&(70-120\log2+16\bar l_1+32\bar l_2+3\bar l_3)\\
\notag
&+\frac{B^{3/2}}{\sqrt2\pi}&&\biggl\{\zeta(-\tfrac12,\tfrac12-\tfrac{3m_\pi^2}{2B})+\zeta(-\tfrac12,\tfrac12) -\int_0^\infty\frac{\dd P}{2\pi}\frac4{1+P^2}\biggl[\zeta\bigl(-\tfrac12,\tfrac12+\tfrac{m_\pi^2(1+P^2)}{2B}\bigr)\\ \notag
&&&
+\frac23\biggl(\frac{m_\pi^2}{2B}\biggr)^{3/2}(1+P^2)^{3/2}\biggr]-\int_0^\infty\frac{\dd P}{2\pi}\frac8{4+P^2}\biggl[\zeta\bigl(-\tfrac12,\tfrac12+\tfrac{m_\pi^2(1+P^2)}{2B}\bigr)\\
\label{FNLOT0}
&&&+\left(\frac{m_\pi^2}{2B}\right)^{3/2}\biggl(\frac23(4+P^2)^{3/2}-3(4+P^2)^{1/2}\biggr)\biggr]\biggr\},
\end{align}
using the notation for the finite counterterms, introduced in section~\ref{sec:renormalization}.

The terms on the first line of eq.~\eqref{FNLOT0} arise from the finite part of the neutral pion contribution~\eqref{FT0wallneutral} and the renormalization of the divergent parts of both eq.~\eqref{FT0wallneutral} and eq.~\eqref{F0wallchargedaux}. The rest of eq.~\eqref{FNLOT0} corresponds to the finite part of the charged pion contribution~\eqref{F0wallchargedaux}. Note that for $B<3m_\pi^2$, the magnetic-field-dependent part of the zero-temperature free energy [in particular the first term on the second line of eq.~\eqref{FNLOT0}] develops an imaginary part. This corresponds to the fact that the domain wall is unstable with respect to charge pion fluctuations unless $B\geq3m_\pi^2$~\cite{Son2008a}. In the opposite corner of the parameter space, $B\gg m_\pi^2$, the free energy is dominated by the charged pion contribution, and eq.~\eqref{FNLOT0} is then well approximated by the very simple expression
\begin{equation}
\left.\frac{\Fa_\text{1,wall}^{T=0}}S\right\rvert_{B\gg m_\pi^2}\simeq\frac{3\log 2}{4\pi^2}m_\pi B.
\label{FT0wallaux}
\end{equation}


\subsubsection{Thermal part}\label{sec:FwallThermal}

Let us now turn to the thermal corrections to the free energy on the domain wall background. While the charged pions turned out to dominate the zero-temperature loop corrections, we expect the situation to be the opposite for thermal corrections. Namely, unlike the neutral pion fluctuations that include the gapless phonon of the CSL (which in the limiting case of the domain wall reduces to the translation zero mode of the wall), the charged pions are gapped, and their contribution is, therefore, expected to be exponentially suppressed at low temperatures.

The thermal part of the free energy of neutral pions is obtained by taking eq.~\eqref{F1wall} and carrying out the Matsubara sum, the result being
\begin{equation}
\frac{\Fa_\text{1,wall}^{T,(\pi^0)}}S=T\sum_\lambda\int\frac{\dd^2\vek p_\perp}{(2\pi)^2}\log\Bigl[1-e^{-\beta\sqrt{\vek p_\perp^2+m_\pi^2(1+\lambda)}}\Bigr].
\end{equation}
The sum over $\lambda$ is then done using eq.~\eqref{lambdasumneutral}. In the contribution of the bound state of the Hamiltonian~\eqref{Hamwall}, $\lambda=-1$, the remaining integral over $\vek p_\perp$ can be evaluated analytically. In the contribution of the continuous spectrum, in turn, the integral over $\vek p_\perp$ can be simplified by introducing spherical coordinates and performing the angular integration. The final result for the thermal free energy due to neutral pions can then be written as
\begin{equation}
\label{FNLOTneutral}
\frac{\Fa_\text{1,wall}^{T,(\pi^0)}}S=-\frac{\zeta(3)T^3}{2\pi}-\frac{m_\pi^2T}{\pi^2} \int_0^\infty P\arctan P \log\Bigl(1-e^{-\beta m_\pi\sqrt{1+P^2}}\Bigr)\,\dd P.
\end{equation}
This expression for $\Fa_\text{1,wall}^{T,(\pi^0)}$ is suitable for numerical evaluation thanks to the exponential convergence of the integral. To get analytic insight in this result, one can represent it as a chosen scale, for instance $T^3$, times a function of the dimensionless parameter $\beta m_\pi$. For $T\ll m_\pi$, the contribution of the continuous spectrum of the domain wall is exponentially suppressed. The thermal free energy is dominated by the gapless surface mode on the domain wall, that is the first term in~\eqref{FNLOTneutral}. On the other hand, $T\gg m_\pi$ can be interpreted as the chiral limit, in which the spectra of neutral pion fluctuations above the CSL state and normal states coincide. Indeed, it can be shown that eq.~\eqref{FNLOTneutral} vanishes in this limit.

Next we turn to the charged pion contribution to the thermal free energy. In order to get a compact expression, we introduce the shorthand notation
\begin{equation}\label{defEpsm}
\epsilon(m,\lambda)\equiv\sqrt{(2m+1)B+m_\pi^2(1+\lambda)}.
\end{equation}
Carrying out the indicated Matsubara sum in eq.~\eqref{F2wall} and subsequently using eq.~\eqref{lambdasumcharged} to do the sum over the eigenvalues $\lambda$, the thermal free energy of charged pion fluctuations can be brought to the form
\begin{equation}
\begin{split}
\frac{\Fa_\text{1,wall}^{T,(\pi^\pm)}}S=\frac{BT}\pi\sum_{m=0}^\infty\biggl\{&\log\left[1-e^{-\beta\epsilon(m,-4)}\right]+\log\left[1-e^{-\beta\epsilon(m,-1)}\right]\\
&-\int_0^\infty\frac{\dd P}{2\pi}\left(\frac4{1+P^2}+\frac8{4+P^2}\right)\log\left[1-e^{-\beta\epsilon(m,P^2)}\right]\biggr\}.
\end{split}
\label{FTwallcharged}
\end{equation}
Note that the instability of the domain wall at any $B<3m_\pi^2$ manifests itself also in this expression. Namely, the lowest-lying charged pion excitation, which is localized in all three dimensions, contributes to the $m=0$ part of the first term in eq.~\eqref{FTwallcharged}, which tends to negative infinity as $B$ approaches $3m_\pi^2$ from above since
\begin{equation}\label{limEps}
\lim_{B\to3m_\pi^2+}\epsilon(0,-4)=0.
\end{equation}
For the same reason, the free energy~\eqref{FTwallcharged} becomes complex for any $B<3m_\pi^2$.


\subsubsection{Asymptotic expansion of the free energy}

The renormalized NLO free energy of the domain wall is given by the combination of the contributions~\eqref{FNLOT0},~\eqref{FNLOTneutral} and~\eqref{FTwallcharged},
\begin{equation}\label{FwallNLO}
\Fa_\text{1,wall}=  \Fa_\text{1,wall}^{T=0}+\Fa_\text{1,wall}^{T,(\pi^0)}+\Fa_\text{1,wall}^{T,(\pi^\pm)}.
\end{equation}
This expression depends on three different scales: $m_\pi$, $T$ and $\sqrt B$. A useful, simple analytic approximation to the NLO free energy exists when $m_\pi$ is much smaller than the other two scales, that is close to the chiral limit. As mentioned above, the zero-temperature free energy is then dominated by the charged pion contribution, whereas the thermal contribution due to neutral pions vanishes in the chiral limit. Furthermore, eq.~\eqref{FTwallcharged} is dominated by the second line and can be simplified by introducing the dimensionless ratio $\gamma\equiv\sqrt{B}/T$ and the auxiliary function~\cite{Agasian2001a}
\begin{equation}
\varphi(\gamma)\equiv\sum_{m=0}^\infty\int_0^\infty\frac{\dd t}{\omega_m(t)\bigl[e^{\gamma\omega_m(t)}-1\bigr]},\qquad
\omega_m(t)\equiv\sqrt{2m+1+t^2}. 
\end{equation}
Altogether, the complete (LO and NLO) free energy of the domain wall in the asymptotic regime $m_\pi \ll T,\sqrt{B}$ can be written in the simple form
\begin{equation}
\frac{\Fa_\text{wall}}S \equiv \frac{\Fa_\text{0,wall}}{S} + \frac{\Fa_\text{1,wall}}{S}\simeq 8m_\pi f_\pi^2-\frac{\mu B}{2\pi}+\frac{3\log 2}{4\pi^2}m_\pi B-\frac{m_\pi T^2}6-\frac{6m_\pi B}{\pi^2}\varphi(\gamma).
\label{Fapprox}
\end{equation}

Remarkably, this approximation to the free energy can be interpreted solely in terms of one-loop renormalization of the pion mass and decay constant~\cite{Agasian2001a},
\begin{align}
\notag
M_\pi^2(B,T)&=m_\pi^2\left[1-\frac{T^2}{24f_\pi^2}-\frac{B\log2}{16\pi^2f_\pi^2}+\frac{B}{2\pi^2f_\pi^2}\varphi(\gamma)\right]+\dotsb,\\
F_\pi^2(B,T)&=f_\pi^2\left[1+\frac{B\log2}{8\pi^2f_\pi^2}-\frac{B}{\pi^2f_\pi^2}\varphi(\gamma)\right]+\dotsb.
\end{align}
Equation~\eqref{Fapprox} can also be used to get some insight into the position of the phase transition from the normal to the CSL phase. This can be located by requiring that $\Fa_\text{wall}=0$. At zero temperature, eq.~\eqref{Fapprox} tells us that unlike at LO, the CSL phase can no longer be the ground state at arbitrary small $\mu$. The threshold value of the chemical potential is $\mu_{\min} = \frac{3\log 2}{2\pi}m_\pi \approx 0.33\,m_\pi$. Moreover, eq.~\eqref{Fapprox} indicates that the CSL phase is  stabilized by thermal corrections. A numerical study of the phase diagram based on the full NLO free energy will be presented in section~\ref{sec:PhaseDiag}, and confirms the above expectations.


\subsection{Next-to-leading-order magnetization}\label{sec:MagWall}

The magnetization of a system can in general be calculated from the free energy as
\begin{equation}\label{defMag}
M=-\frac{\partial \Fa}{\partial B}.
\end{equation}
At LO, the magnetization of the domain wall per unit surface follows at once from~\eqref{Ewall},
\begin{equation}\label{MDWLO}
\frac{M_\text{0,wall}}{S} = \frac{\mu}{2\pi},
\end{equation}
independently of the magnetic field~\cite{Son2004a}. To find the NLO correction to the magnetization, we need to differentiate eqs.~\eqref{FNLOT0} and~\eqref{FTwallcharged} with respect to the magnetic field.\footnote{Recall that these expressions give the difference of free energies of the domain wall and the normal phase. However, in the normal phase, both the free energy and the magnetization are distributed uniformly in the bulk. What we calculate here is the excess magnetization associated with the domain wall. Note also that the neutral pion contribution to the NLO free energy~\eqref{FNLOTneutral} is independent of the magnetic field and thus does not contribute to the magnetization.} To that end, we use the identity
\begin{equation}\label{HurwitzDer}
\frac{\partial \zeta(s,q)}{\partial q} = -s \,\zeta(s+1,q),
\end{equation}
which follows directly from the definition~\eqref{HurwitzDef}. One then finds
\begin{align}\label{magWallT0}
\frac{M_\text{1,wall}^{T=0}}{S}= -\frac{\sqrt{B}}{\sqrt2\pi}\biggl\{&\frac{3}{2}\zeta(-\tfrac12,\tfrac12-\tfrac{3m_\pi^2}{2B}) + \frac{3m_\pi^2}{4B}\zeta(\tfrac12,\tfrac12-\tfrac{3m_\pi^2}{2B}) + \frac{3}{2}\zeta(-\tfrac12,\tfrac12)\\ \notag
&-\int_0^\infty\frac{\dd P}{2\pi}\left(\frac4{1+P^2}+\frac8{4+P^2}\right)\biggl[\frac{3}{2}\zeta\bigl(-\tfrac12,\tfrac12+\tfrac{m_\pi^2(1+P^2)}{2B}\bigr)\\ \notag
&-\frac{m_\pi^2 (1+P^2)}{4B}\zeta\bigl(\tfrac12,\tfrac12+\tfrac{m_\pi^2(1+P^2)}{2B}\bigr)\biggr]\biggr\}
\end{align}
for the NLO contribution to the magnetization of the domain wall per unit surface at zero temperature, and
\begin{align}\label{magWallT}
\frac{M_\text{1,wall}^{T}}{S}=-\frac{1}{\pi} \sum_{m=0}^\infty  \Biggl\{ &T \log\left[1-e^{-\beta\epsilon(m,-4)}\right]+ \frac{(2m+1)B}{2 \epsilon(m,-4) \left[e^{\beta\epsilon(m,-4)}-1\right]}\\ \notag
 & + T \log\left[1-e^{-\beta\epsilon(m,-1)}\right] + \frac{(2m+1)B}{2 \epsilon(m,-1) \left[e^{\beta\epsilon(m,-1)}-1\right]}\\ \notag
&-\int_0^\infty\frac{\dd P}{2\pi}\left(\frac4{1+P^2}+\frac8{4+P^2}\right)\biggl(T \log\left[1-e^{-\beta\epsilon(m,P^2)}\right] \\ \notag
&+ \frac{(2m+1)B}{2 \epsilon(m,P^2) \left[e^{\beta\epsilon(m,P^2)}-1\right]} \biggr) \biggr\}
\end{align}
for the additional thermal contribution. The zero-temperature contribution to magnetization diverges in the limit $B\to 3 m_\pi^2$ thanks to
\begin{equation}
\lim_{x\to0+}\zeta(1/2,x)=+\infty.
\label{limZeta}
\end{equation}
The thermal contribution likewise diverges in the same limit due to the property~\eqref{limEps}.
 

\section{Chiral soliton lattice at next-to-leading order}
\label{sec:CSL}

\subsection{General strategy}

In case of the more general CSL solution, we have to follow the same steps to evaluate the neutral and charged pion contributions to the one-loop free energy, given by eqs.~\eqref{F1CSL} and~\eqref{F2CSL}. Before we split up the calculation into the zero-temperature and thermal parts, it is, however, worth outlining the strategy for evaluation of the sums over the eigenvalues $\lambda$ of the effective Hamiltonians~\eqref{HamCSL} and~\eqref{HamCSLcharged}, similarly to the domain wall case.


\subsubsection{Neutral pions}

Let us illustrate the strategy in detail on the simple case of the neutral pion contribution. Upon explicitly subtracting the corresponding contribution to the free energy of the normal phase, eq.~\eqref{F1CSL} can be written as
\begin{equation}
\beta\Fa_\text{1,CSL}^{\text{1-loop},(\pi^0)}={}\frac12\sum_{n,\vek p_\perp,P}\left\{\log\left[\Omega_0^2 + \lambda(P) -2k^2\right] - \log\left(\Omega_0^2+P^2\right)\right\},
\label{F1loopCSL}
\end{equation}
where $P \equiv p_z k/m_\pi$ is the dimensionless quasi-momentum in the direction of the magnetic field, $\lambda(P)$ is the corresponding eigenvalue of the operator~\eqref{HamCSL}, and
\begin{equation}
\Omega_0^2\equiv\frac{k^2}{m_\pi^2}(\omega_n^2+\vek p_\perp^2+m_\pi^2).
\label{Omega0def}
\end{equation}
Now that the background is periodic and the spectrum can accordingly be characterized by a momentum variable, the discrete sum over $\vek p_\perp$ and $P$ can be converted into a continuous integral through the usual replacement
\begin{equation}
\sum_{\vek p_\perp,P}\to V\frac{m_\pi}k\int\frac{\dd^{d-1}\vek p_\perp}{(2\pi)^{d-1}}\frac{\dd P}{2\pi},
\end{equation}
The factor $m_\pi/k$ relates the dimensionless variable $P$ to the physical quasi-momentum $p_z$ and $d=3-2\epsilon$ will again be used for dimensional regularization.

The integral over $P$ cannot be performed directly because an explicit expression for the dimensionless ``energy'' $\lambda(P)$ is not available. What is known is the group velocity~\cite{Li2000a}
\begin{equation}
\frac{\dd\lambda}{\dd P}=\frac{2\sqrt{(\lambda-1)(\lambda-k^2)(\lambda-1-k^2)}}{k^2+\frac{E(k)}{K(k)}-\lambda} \equiv f_1(\lambda).
\label{DoSneutral}
\end{equation}
This is sufficient to convert the integral over $P$ into one over $\lambda$. The latter requires some care since the spectrum of neutral pion fluctuations of the CSL solution spans two bands covering the intervals
\begin{equation}
k^2\leq\lambda\leq 1\qquad\text{and}\qquad
1+k^2\leq\lambda.
\label{bandsneutral}
\end{equation}
(See appendix~\ref{app:spectra} for more details on the spectrum of the corresponding Lam\'{e} Hamiltonian.) Remembering to add an overall factor of $2$ to account for the double degeneracy of energy levels in the continuous spectrum, we then arrive at the following expression for the neutral pion contribution to the one-loop free energy of the CSL state,\footnote{The one-dimensional integral without specified integration bounds is to be understood as an integral over the whole real axis; this notation will be used throughout the text.}
\begin{align}
\label{monster}
\frac{\beta\Fa_\text{1,CSL}^{\text{1-loop},(\pi^0)}}V=&\frac12\frac{m_\pi}k \left(\frac{e^{\gamma_\text{E}}\Lambda_\text{RG}^2}{4\pi}\right)^\epsilon \sum_n\int\frac{\dd^{d-1}\vek p_\perp}{(2\pi)^{d-1}}\biggl\{\\
&\int \frac{\dd\lambda}{\pi f_1(\lambda)}\log(\Omega_0^2+\lambda-2k^2) \left[\chi_{(k^2,\,1)}-\chi_{(1+k^2,\,\infty)}\right]
\notag
-\int \frac{\dd P}{2\pi}\log(\Omega_0^2+P^2)\biggr\}.
\end{align}
Here $\chi_I$ is the characteristic function of an interval $I$. The different signs in front of the two characteristic functions in eq.~\eqref{monster} are related to the fact that $f_1(\lambda)$ arising from the change of variables from $P$ to $\lambda$ has a different sign in the two energy bands.

Remarkably, the difference of the two integrals in the curly brackets in eq.~\eqref{monster} can be evaluated analytically. By using an alternative approach to the calculation of the functional determinant of $\Da_\text{CSL}^{(\pi^0)}$, based on the Gelfand-Yaglom theorem~\cite{Dunne2008a}, one can show that this difference of integrals equals
\begin{equation}
Z(u,k)+\cs(u,k)\dn(u,k)-\Omega_0,\\
\label{Gelfand}
\end{equation}
where $Z(u,k)$ is the Jacobi zeta function and $u$ is given implicitly by solving the condition
\begin{equation}
\cs^2(u,k)=\Omega_0^2-k^2=\frac{k^2}{m_\pi^2}(\omega_n^2+\vek p_\perp^2).
\label{csak}
\end{equation}
See appendix~\ref{app:Gelfand} for some details of the derivation of eq.~\eqref{Gelfand}.

Although it might be tempting to replace the complicated integrals in eq.~\eqref{monster} with the compact expression~\eqref{Gelfand}, it is in fact the former expression that is more useful. Namely, the fact that the Matsubara frequency $\omega_n$ and the transverse momentum $\vek p_\perp$ only enter eq.~\eqref{monster} through the logarithms therein makes it possible to carry out the Matsubara sum and the integral over $\vek p_\perp$ in a closed form, without having to worry about the band spectrum of CSL.


\subsubsection{Charged pions}

The strategy for the charged pion case is analogous, albeit technically more involved. Again, the free energy of the normal phase is explicitly subtracted from the contribution of eq.~\eqref{F2CSL} and the sum over the dimensionless quasi-momenta $P$ is turned into a continuous integral in the standard way,
\begin{align}\notag
\frac{\beta\Fa^{\text{1-loop},(\pi^\pm)}_\text{1,CSL}}{V}=\frac{m_\pi B}{2\pi k}\sum_{n,m}\int \frac{\dd P}{2\pi} \biggl\{{}&\log\left[\omega_n^2 + (2m+1)B + \frac{m_\pi^2}{k^2}(\lambda(P)-4-k^2) \right] \\
&- \log\left[\omega_n^2 + (2m+1)B + \frac{m_\pi^2}{k^2}(P^2+k^2) \right]\biggr\}.\label{F1loopCharged}
\end{align}
Like in the neutral pion case, an expression for the dimensionless energy $\lambda(P)$ is not known explicitly, but the group velocity can be inferred, e.g., from ref.~\cite{Maier2008a},
\begin{align}\label{DoScharged}
\frac{\dd\lambda}{\dd P}=\frac{2\sqrt{\Pi_{j=1}^5(\lambda-\lambda_j)}}{ 2 - (\lambda - 4 k^2) (\lambda -2 - k^2) + 3 (\lambda - 2  -2  k^2) \frac{E(k)}{K(k)}} \equiv f_2 (\lambda),
\end{align}
where 
\begin{equation}
\begin{gathered}
\lambda_1=2 \left(1+  k^2- \sqrt{1-k^2+k^4}\right),\qquad \lambda_2 =1+k^2,\qquad \lambda_3 =1+4k^2,\\
\lambda_4=4+k^2,\qquad \lambda_5=2\left(1+k^2+\sqrt{1-k^2+k^4}\right).
\end{gathered}
\label{lambdasCharged}
\end{equation}
The spectrum of the operator~\eqref{HamCSLcharged} (the $n=2$ Lam\'{e} Hamiltonian) consists of three bands spanning the intervals
\begin{equation}
\lambda_1\leq \lambda \leq \lambda_2,\qquad\lambda_3\leq \lambda \leq \lambda_4,\qquad \lambda_5\leq \lambda;
\end{equation}
see figure~\ref{fig:specLame} in appendix~\ref{app:spectra} for a visualization of this band spectrum.

As pointed out already in ref.~\cite{Brauner2017a}, the bottom of the lowest Landau level can reach zero for sufficiently strong magnetic fields, which indicates an instability of CSL under Bose-Einstein condensation (BEC) of charged pions. The critical magnetic field $B_\text{BEC}$ for this BEC can be inferred from eq.~\eqref{F1loopCharged}. Namely, the argument of the first logarithm therein for $m=n=0$ and $\lambda=\lambda_1$ reaches zero if
\begin{equation}\label{BBEC}
B_\text{BEC} = \frac{m_\pi^2}{k^2}\left(4 + k^2 - \lambda_1 \right).
\end{equation}
Here $k=k(\mu,B_\text{BEC})$ is fixed by the condition~$\eqref{condk}$. Hence, eq.~\eqref{BBEC} determines the critical field $B_\text{BEC}$ implicitly in terms of the baryon chemical potential. Further analytic insight is possible in certain limiting regimes. First, in the limit $k\to 1$, $B_\text{BEC}\to 3m_\pi^2$, which is consistent with the observed instability of the domain wall spectrum. Second, for $B\gg m_\pi^2$ (i.e., $k \ll 1 $), the condition~$\eqref{condk}$ simplifies to~\cite{Brauner2017b}
\begin{equation}\label{condkLargeB}
k \approx \frac{8\pi^2f_\pi^2}{\mu}\,\frac{m_\pi}{B}.
\end{equation}
Using this in eq.~\eqref{BBEC} and discarding terms that are subleading in the limit $k \ll 1 $, one can then easily solve for the critical field,
\begin{equation}\label{BBECchiral}
B_{\text{BEC}} \approx \frac{16 \pi^4 f_\pi^4}{\mu^2} \qquad(\text{for } B\gg m_\pi^2).
\end{equation}
Note that the regime $k\ll1$ is equivalent to the chiral limit, $m_\pi \to 0$. It was shown in ref.~\cite{Evans:2022hwr} that in the chiral limit, a nontrivial configuration of charged pion fields is indeed energetically preferable above the critical magnetic field~\eqref{BBECchiral}.


\subsection{One-loop free energy: zero-temperature part}

We now start with the zero-temperature part of the one-loop free energy, which can be extracted from eqs.~\eqref{monster} and~\eqref{F1loopCharged} by converting the Matsubara sum into a continuous frequency integral. This part of the CSL free energy requires renormalization, which is accomplished by adding extra contributions arising from the counterterms in eq.~\eqref{Lct}. Since the free energy of the CSL state scales with volume, it is sufficient to evaluate the spatial averages of the counterterm operators in eq.~\eqref{Lct}. These are given by identities similar to eq.~\eqref{CTwall},
\begin{align}
\notag
\langle[\phi_0'(z)]^4\rangle={}&\frac{16m_\pi^4}3\biggl[-\frac1{k^4}+\frac1{k^2} +\left(\frac4{k^4}-\frac2{k^2}\right)\frac{E(k)}{K(k)}\biggr],\\
\langle[\phi_0'(z)]^2\cos\phi_0(z)\rangle={}&\frac{8m_\pi^2}3\biggl[\frac1{k^4}-\frac1{k^2} +\left(-\frac1{k^4}+\frac1{2k^2}\right)\frac{E(k)}{K(k)}\biggr],\\
\notag
\langle\cos^2\phi_0(z)-1\rangle={}&\frac8{3k^4}-\frac8{3k^2}+\left(-\frac8{3k^4}+\frac4{3k^2}\right)\frac{E(k)}{K(k)},
\end{align}
where the CSL solution~\eqref{solCSL} has been used. The corresponding counterterm free energy then reads
\begin{equation}
\frac{\Fa_\text{1,CSL}^\text{c.t.}}V=-\ell_1\langle[\phi_0'(z)]^4\rangle+\ell_2m_\pi^2\langle[\phi_0'(z)]^2\cos\phi_0(z)\rangle -\ell_3m_\pi^4\langle\cos^2\phi_0(z)-1\rangle.
\label{FCTCSL}
\end{equation}


\subsubsection{Neutral pion contribution}

Let us now turn to the zero-temperature part of eq.~\eqref{monster}. Analogously to eq.~\eqref{FT0NeutralWall1}, we combine the frequency $\omega$ with the transverse momentum $\vek p_\perp$ into a single transverse energy-momentum variable $P_\perp$. The integral over $P_\perp$ can be done in a closed form. This brings eq.~\eqref{monster} into a form where only a one-dimensional integral remains to be done,
\begin{align}\notag
\frac{\Fa_\text{1,CSL}^{T=0,(\pi^0)}}V= -\frac{\left(e^{\gamma_\text{E}}\Lambda_\text{RG}^2\right)^\epsilon m_\pi^{d+1}}{2k}\frac{\Gamma(-\frac d2)}{(4\pi)^{3/2}}\biggl\{{}&\int\frac{\dd\lambda}{\pi f_1(\lambda)}\left(\frac\lambda{k^2}-1\right)^{d/2} \left[\chi_{(k^2,\,1)}-\chi_{(1+k^2,\,\infty)}\right] \\
&-\int \frac{\dd P}{2\pi}\left(\frac{P^2}{k^2}+1\right)^{d/2}\biggr\}.
\label{monster2}
\end{align}
The integral over the lower energy band arising from the term proportional to $\chi_{(k^2,\,1)}$ is finite and can be evaluated in a closed form. As for the remaining integrals, we need to extract their divergent parts, which is best done using a momentum variable instead of $\lambda$. Thus, in the integral over the upper energy band, we substitute $\lambda=1+k^2+P^2$. This brings eq.~\eqref{monster2} to the form
\begin{align}\label{monster3}
\frac{\Fa_\text{1,CSL}^{T=0,(\pi^0)}}V=&-\frac{\left(e^{\gamma_\text{E}}\Lambda_\text{RG}^2\right)^\epsilon m_\pi^{d+1}}{2k}\frac{\Gamma(-\frac d2)}{(4\pi)^{3/2}}\biggl\{\int_{k^2}^1\frac{\dd\lambda}{\pi f_1(\lambda)} \left(\frac\lambda{k^2}-1\right)^{d/2}\\
&+\frac1{k^d}\int\frac{\dd P}{2\pi}\biggl[(P^2+1)^{d/2}\frac{P^2+1-\frac{E(k)}{K(k)}}{\sqrt{(P^2+1)(P^2+k^2)}}-(P^2+k^2)^{d/2}\biggr]\biggr\}.
\notag
\end{align}
The next step is to expand the integrand on the second line in powers of $P^2+1$. In this way, the divergent part of eq.~\eqref{monster3} can be identified and combined with the appropriate counterterm. In particular, one third of the $\ell_3$ operator in eq.~\eqref{FCTCSL} is necessary to cancel the divergence of the integral. For the sake of bookkeeping, we choose to add solely the pole part of the counterterm here; all the finite parts of the counterterm Lagrangian will be added to the charged pion contribution~\eqref{FT0CSLCharged} below. 

In the finite part of eq.~\eqref{monster3}, $d=3$ can be set and all the remaining integrals can be evaluated analytically. A closed, analytic expression for the renormalized contribution of neutral pions to the zero-temperature part of the free energy of the CSL state follows,
\begin{align}\notag
\frac{\Fa_\text{1,CSL}^{T=0,(\pi^0)}}V=\frac{m_\pi^4}{24\pi^2k^4}\Biggl\{&-\frac76(1-k^2)+\frac53\left(1-\frac{k^2}2\right)\frac{E(k)}{K(k)}+\left[-\frac32+\frac{3k^2}4+\frac{E(k)}{K(k)}\right]\sqrt{1-k^2}\\
& +\left[2-2k^2+\frac{3k^4}4+\left(-2+k^2\right)\frac{E(k)}{K(k)}\right]\log\frac{1+\sqrt{1-k^2}}{k}\Biggr\}.\label{FT0CSLNeutral}
\end{align}


\subsubsection{Charged pion contribution}

The temperature-independent part of~\eqref{F1loopCharged} can be extracted using the replacement~\eqref{repOmega},
\begin{align}
&\frac{\Fa^{T=0,(\pi^\pm)}_\text{1,CSL}}{V}=\frac{m_\pi B}{2\pi k} \left(\frac{e^{\gamma_\text{E}}\Lambda_\text{RG}^2}{4\pi}\right)^\epsilon \int\frac{\dd^{d-2}\omega}{(2\pi)^{d-2}}\sum_{m}\int \frac{\dd P}{2\pi}\biggl\{ 
\\
&\log\biggl[\omega^2 + (2m+1)B + \frac{m_\pi^2}{k^2}(\lambda(P)-4-k^2) \biggr] 
- \log\biggl[\omega^2 + (2m+1)B + \frac{m_\pi^2}{k^2}(P^2+k^2) \biggr]\biggr\} .\notag
\end{align}
The integral over frequency can be easily evaluated and the sum over Landau levels can be turned into the Hurwitz $\zeta$-function as in the case of the domain wall background. The integral over $P$ can then be turned into one over $\lambda$ using~\eqref{DoScharged}. As a result,
\begin{align}\notag
&\frac{\Fa^{T=0,(\pi^\pm)}_\text{1,CSL}}{V}=-\frac{m_\pi \left(e^{\gamma_\text{E}}\Lambda_\text{RG}^2\right)^\epsilon}{k}\frac{\Gamma(1-\frac{d}{2})}{(4\pi)^{3/2}}(2B)^{d/2} \Biggl\{\\ \notag
&\int \frac{\dd \lambda }{\pi f_2 (\lambda)} \, 
\zeta\bigl(1-\tfrac{d}{2},\tfrac{1}{2}+\tfrac{m_\pi^2}{2Bk^2}[\lambda-4-k^2] \bigr) \left[-\chi_{(\lambda_1,\lambda_2)}+\chi_{(\lambda_3,\lambda_4)}-\chi_{(\lambda_5,\infty)}\right] \\
&- \int \frac{\dd P }{2\pi} \zeta\bigl(1-\tfrac{d}{2},\tfrac{1}{2}+\tfrac{m_\pi^2}{2Bk^2}[P^2+k^2] \bigr) \Biggr\},
\end{align}
where again the signs in front of the different characteristic functions arise from the signs of $f_2(\lambda)$ in different energy bands. The divergent part of this expression comes from the integration over the third energy band and the contribution of the normal phase. In order to combine these two contributions into a single integral, the substitution $\lambda=P^2+\lambda_5$ can be introduced in the former integral. The resulting integrand can be expanded in powers of $P^2+k^2$ with the help of eq.~\eqref{Hurwitz2} in order to find the asymptotic behavior for large $P$. The divergence for $d \to 3$ is canceled by the remaining piece of the counterterm Lagrangian~\eqref{FCTCSL}. Recalling that we are now to include the entire finite part of the counterterm Lagrangian, a careful evaluation of all the contributions leads to
\begin{align}\notag
\frac{\Fa_\text{1,CSL}^{T=0,(\pi^\pm)}}V=\frac{m_\pi^4}{24\pi^2 k^4}\Biggl\{& 15 - 14k^2+ 3 k^4 +\left(\frac{25}{2} k^2 - 19\right)\frac{E(k)}{K(k)} - \left(4 + 12 \frac{E(k)}{K(k)}\right)\sqrt{1-k^2+k^4}\\ \notag
&+(1-k^2)\left(\frac{4}{3}\bar{l}_1+\frac{8}{3}\bar{l}_2 + \bar{l}_3 \right) -\left(1-\frac{k^2}{2}\right)\frac{E(k)}{K(k)}\left(\frac{16}{3}\bar{l}_1+\frac{32}{3}\bar{l}_2+\bar{l}_3\right)\Biggr\}\\ \notag
+ \frac{m_\pi B^{3/2}}{\sqrt{2}k\pi}\Biggl\{&\int \frac{\dd \lambda }{\pi f_2(\lambda)} \, 
\zeta\bigl(-\tfrac{1}{2},\tfrac{1}{2}+\tfrac{m_\pi^2}{2Bk^2}[\lambda-4-k^2] \bigr) \left[-\chi_{(\lambda_1,\lambda_2)}+\chi_{(\lambda_3,\lambda_4)}\right] \\ \notag
&+ \int_0^\infty \frac{\dd P }{\pi}\biggl[ -\frac{2P}{f_2(P^2+\lambda_5)}\zeta\bigl(-\tfrac{1}{2},\tfrac{1}{2}+\tfrac{m_\pi^2}{2Bk^2}[P^2+\lambda_5-4-k^2] \bigr)\\ \label{FT0CSLCharged}
& -\zeta\bigl(-\tfrac{1}{2},\tfrac{1}{2}+\tfrac{m_\pi^2}{2Bk^2}[P^2+k^2] \bigr)\\ \notag
&-\left(\frac{m_\pi^2}{2Bk^2}\right)^{3/2}\left[C_1(k)(P^2+k^2)^{1/2}+C_2(k)(P^2+k^2)^{-1/2}\right]
 \biggr]\Biggr\};
\end{align}
the auxiliary functions $C_1$ and $C_2$ of the elliptic modulus $k$ are defined by
\begin{align}\notag
C_1(k) &\equiv\frac{2}{3} \left(1+k^2+3\frac{E(k)}{K(k)}-2 \sqrt{1-k^2+k^4}\right),\\ 
C_2(k) &\equiv2-\frac{7}{3}k^2 -\frac{1}{3}k^4 +\frac{2}{3}k^2\sqrt{1-k^2+k^4} - \left(6-2k^2\right)\frac{E(k)}{K(k)}.
\end{align}
Note that the Hurwitz $\zeta$-function on the third line of eq.~\eqref{FT0CSLCharged} can pick up an imaginary part if its second argument turns negative, which, in turn, happens if the magnetic field is larger than the critical value~\eqref{BBEC} for BEC of charged pions. However, the value of eq.~\eqref{FT0CSLCharged} remains finite at this critical magnetic field. 

The full zero-temperature NLO free energy of CSL is given by the sum of eqs.~\eqref{FT0CSLNeutral} and~\eqref{FT0CSLCharged}. It tends to zero in the limit $k\to 1$ as expected since $k=1$ corresponds to the domain wall case where only the free energy per unit \emph{surface} is nonzero. However, let us inspect the asymptotic behavior for $k\to 1$ in more detail since this can give us a nontrivial check of the correctness of our result. To this end, recall that~\cite{Abramowitz1972a}
\begin{equation}
\lim_{k\to 1} \left[K(k)-\frac{1}{2}\log\frac{16}{1-k^2}\right] = 0,
\end{equation}
hence $K(k) \propto \log (1-k)$ as $k\to 1$. Consequently, the terms in the NLO free energy proportional to $1/K(k)$ are dominant for $k\to 1$. To extract the asymptotic behavior for $k\to1$, one can take the limit $k\to1$ everywhere except for the very $1/K(k)$ terms. In doing so, the first two energy bands in the charged pion spectrum shrink to points, yet the third line of eq.~\eqref{FT0CSLCharged} remains nonzero. This is because after setting $k=1$ wherever possible, the integrals over the first and second energy bands can be reduced to the form
\begin{equation}\label{kto1Integral}
\int_0^c \frac{dx}{\sqrt{x(c-x)}} = \pi,
\end{equation}
where $c$ is a constant that tends to zero as $k\to1$. Keeping this subtlety in mind, one recovers the domain wall result of eq.~\eqref{FNLOT0},
\begin{equation}\label{F0CSLkto1}
\left(\frac{\Fa^{T=0,(\pi^0)}_\text{1,CSL}}{V}+\frac{\Fa^{T=0,(\pi^\pm)}_\text{1,CSL}}{V}\right)\Biggr|_{k\to 1}\simeq \frac{m_\pi}{2 K(k)} \frac{\Fa^{T=0}_\text{1,wall}}{S}.
\end{equation}
The normalization of the right-hand side agrees  with the fact that the period of the CSL solution is $2 k K(k)/m_\pi$.


\subsection{One-loop free energy: thermal part}

The thermal part of the free energy is in a sense more straightforward to evaluate since it does not require renormalization. On the other hand, the presence of the Matsubara sum makes the analytic expressions necessarily more involved, especially in combination with the sum over Landau levels in the part coming from charged pions.


\subsubsection{Neutral pion contribution}

The neutral pion contribution is obtained from eq.~\eqref{monster} by taking the limit $d\to3$ and keeping only the thermal part of the Matsubara sum. The remaining integration over transverse momenta can be carried out analytically using the formula
\begin{align}\label{polylog}
\frac T2\sum_n\int\frac{\dd^2\vek p_\perp}{(2\pi)^2}\log(\omega_n^2+\vek p_\perp^2+M^2)
\to & \,T\int\frac{\dd^2\vek p_\perp}{(2\pi)^2}\log\Bigl(1-e^{-\beta\sqrt{\vek p_\perp^2+M^2}}\Bigr)\\
& =-\frac1{2\pi}\bigl[T^3\Li_3\bigl(e^{-\beta M}\bigr)+T^2M\Li_2\bigl(e^{-\beta M}\bigr)\bigr],
\notag
\end{align}
where $\Li_n(z)$ is the polylogarithm, and the arrow indicates dropping the zero-temperature part of the Matsubara sum.

Armed with eq.~\eqref{polylog}, we can write down the thermal part of the neutral pion free energy~\eqref{monster} in terms of a one-dimensional, exponentially convergent integral. It makes the resulting expressions somewhat simpler to trade the variable $\lambda$ for a new variable $t$. In the integral over $k^2\leq\lambda\leq1$, the natural choice of the substitution is $\lambda=k^2+k'^2t^2$, where $k'^2 \equiv 1-k^2$ is the complementary elliptic modulus. In the integral over $1+k^2\leq\lambda<\infty$, either the same substitution, or alternatively $\lambda=1+k^2+k^2t^2$, can be used. The most compact expression for the thermal free energy we can thus get is
\begin{align}\notag
\frac{\Fa_\text{1,CSL}^{T,(\pi^0)}}{VT^4}=&-\frac x{2\pi^2k}\int\dd t\,\frac{k'^2t^2-\frac{E(k)}{K(k)}}{\sqrt{(1-t^2)(1-k'^2t^2)}}\bigl[\chi_{(\frac1{k'},\infty)}-\chi_{(0,1)}\bigr]\biggl[\Li_3\bigl(e^{_-\frac{k'}kxt}\bigr)+\frac{k'}kxt\Li_2\bigl(e^{_-\frac{k'}kxt}\bigr)\biggr]\\
&-\frac1{2\pi^2}\int_0^\infty\dd t\,t^2\log\Bigl(1-e^{-\sqrt{t^2+x^2}}\Bigr),
\label{FTCSLNeutral}
\end{align}
where $x\equiv m_\pi/T$. The thermal free energy can also be cast in an entirely different form by trading the polylogarithms for elliptic integrals via integration by parts. Using the fact that the first factor under the integral on the first line of eq.~\eqref{FTCSLNeutral} can be integrated in a closed form, we can differentiate the combination of the polylogarithms, arriving at
\begin{align}\notag
\frac{\Fa_\text{1,CSL}^{T,(\pi^0)}}{VT^4}=&-\frac x{4\pi kK(k)}\biggl[\Li_3\bigl(e^{_-\frac{k'}kx}\bigr)+\frac{k'}kx\Li_2\bigl(e^{_-\frac{k'}kx}\bigr)\biggr]-\frac1{2\pi^2}\int_0^\infty\dd t\,t^2\log\Bigl(1-e^{-\sqrt{t^2+x^2}}\Bigr)\\ \notag
&-\frac{x^3}{2\pi^2k}\int_0^\infty\dd t\,t\log\Bigl(1-e^{-\frac xk\sqrt{1+k^2t^2}}\Bigr)\biggl\{\biggl[\frac{E(k)}{K(k)}-1\biggr]F(\arctan t,k')\\ \notag
&\hspace{17 em}+E(\arctan t,k')-t\sqrt{\frac{1+k^2t^2}{1+t^2}}\biggr\}\\
&-\frac{k'^2x^3}{2\pi^2k^3}\int_0^1\dd t\,t\log\Bigl(1-e^{-\frac{k'}kxt}\Bigr)\biggl\{\biggl[1-\frac{E(k)}{K(k)}\biggr]F(\arcsin t,k')-E(\arcsin t,k')\biggr\},
\label{FTCSL2}
\end{align}
where $F(u,k)$ and $E(u,k)$ are the elliptic integrals of the first and second kind, respectively, and the very first contribution comes from the surface term of the integral over $0\leq t\leq1$.

The expression~\eqref{FTCSL2} for the thermal free energy is a suitable starting point for taking various analytical limits. 
First of all, let us again inspect the limit $k\to1$ as a consistency check. Following the steps outlined above for the zero-temperature part of the CSL free energy, we find
\begin{equation}\label{FTCSLkto1}
\frac{\Fa_\text{1,CSL}^{T,(\pi^0)}}{V}\Biggr|_{k\to 1}\simeq \frac{m_\pi}{2 K(k)}\frac{\Fa^{T,(\pi^0)}_\text{1,wall}}{S},
\end{equation}
in agreement with the domain wall result~\eqref{FNLOTneutral}.

Next we look at the low-temperature regime, $T\ll m_\pi$ (or $x\gg1$). In this limit, all parts of eq.~\eqref{FTCSL2} but the last line are exponentially suppressed. In the last integral over $0\leq t\leq1$ it is only small values of $t$ that matter. We can therefore expand everything but the logarithm in powers of $t$, work out the integral in a closed form, and eventually expand the result in $x$. The final result reads
\begin{equation}
\frac{\Fa_\text{1,CSL}^{T,(\pi^0)}}V=-\frac{\pi^2T^4}{90}\frac{E(k)}{k'K(k)}-\frac{2\pi^4T^6}{945m_\pi^2}\frac{k^2}{k'^3}\biggl[(1+k'^2)\frac{E(k)}{K(k)}-2k'^2\biggr]+\mathcal O(T^8).
\end{equation}
The leading, $\mathcal O(T^4)$ term is just the free energy of a gas of free phonons with linear dispersion relation and phase velocity calculated in ref.~\cite{Brauner2017a}. The next-to-leading, $\mathcal O(T^6)$ term represents the leading correction due to the nonlinearity of the phonon dispersion relation.


\subsubsection{Charged pion contribution}

The temperature-dependent part of eq.~\eqref{F1loopCharged} is extracted using the replacement
\begin{equation}
\frac{T}{2}\sum_n \log(\omega_n^2+\epsilon^2)\to T\log\bigl(1-e^{-\beta\epsilon}\bigr).
\end{equation}
The momentum integral over the energy bands of the CSL spectrum can be converted into one over the energy eigenvalue $\lambda$ using eq.~\eqref{DoScharged}. This leads to
\begin{align} \notag
\frac{\Fa_\text{1,CSL}^{T,(\pi^\pm)}}{V}=
\frac{m_\pi B T}{k\pi} \sum_{m=0}^\infty \Biggl\{&\int \frac{\dd \lambda }{\pi f_2(\lambda)} \log\left[1-e^{-\beta \epsilon_{\text{CSL}}(m,\lambda)}\right]
 \left[-\chi_{(\lambda_1,\lambda_2)}+\chi_{(\lambda_3,\lambda_4)}-\chi_{(\lambda_5,\infty)}\right] \\
&- \int \frac{\dd P }{2\pi} \log\left[1-e^{-\beta \epsilon_0(m,P)}\right] \Biggr\}, \label{FTCSLCharged}
\end{align}
with the shorthand notation
\begin{equation}
\begin{split}
\epsilon_{\text{CSL}}(m,\lambda)&\equiv \sqrt{(2m+1)B+\frac{m_\pi^2}{k^2} (\lambda-4-k^2)},\\
\epsilon_0(m,P)&\equiv \sqrt{(2m+1)B+\frac{m_\pi^2}{k^2} (P^2+k^2)}.
\end{split}
\label{defEpsB}
\end{equation}
It is straightforward to show that the asymptotic behavior of the thermal free energy of charged pions satisfies a relation analogous to eq.~\eqref{FTCSLkto1}, consistent with the domain wall result~\eqref{FTwallcharged}. Note also that $\epsilon_\text{CSL}(0,\lambda_1)$ tends to zero for $B\to B_{\text{BEC}}$. Hence, the thermal free energy~\eqref{FTCSLCharged} tends to negative infinity in this limit.

Altogether, the complete (LO and NLO) renormalized free energy of the CSL is given by a combination of eqs.~\eqref{ECSL}, \eqref{FT0CSLNeutral}, \eqref{FT0CSLCharged}, \eqref{FTCSLNeutral} and~\eqref{FTCSLCharged},
\begin{equation}\label{FCSLfull}
\Fa_\text{CSL} = \Fa_\text{0,CSL} + \Fa_\text{1,CSL}^{T=0,(\pi^0)}+ \Fa_\text{1,CSL}^{T=0,(\pi^\pm)}+\Fa_\text{1,CSL}^{T,(\pi^0)}+\Fa_\text{1,CSL}^{T,(\pi^\pm)}.
\end{equation}


\subsection{Next-to-leading-order magnetization}\label{sec:MagCSL}

The magnetization of the CSL is in principle simply determined by eq.~\eqref{FCSLfull} in combination with eq.~\eqref{defMag}. There are, however, two subtleties to keep in mind. First, when taking a derivative of the free energy with respect to the magnetic field, only the explicit dependence on $B$ is to be taken into account. The implicit dependence on $B$ through the elliptic modulus $k$ can be disregarded on the LO CSL state thanks to the chain rule. (This is a variation on the Feynman-Hellmann theorem known in quantum mechanics.) Second, one must keep in mind that eq.~\eqref{FCSLfull} represents the \emph{difference} of free energy densities of the CSL and normal states. In order to recover the magnetization of CSL alone, one must add to $-\de\Fa_{\text{CSL}}/\de B$ the magnetization of the QCD vacuum.

At LO, the magnetization of the QCD vacuum vanishes, and the magnetization of CSL then descends directly from eq.~\eqref{ECSL},
\begin{equation}\label{MCSLLO}
\frac{M_{\text{0,CSL}}}{V}=\frac{m_\pi\mu}{4\pi kK(k)}.
\end{equation}
Note that the LO magnetization of one period of the CSL configuration per unit transverse area is equal to the LO domain wall surface magnetization~\eqref{MDWLO}.


\subsubsection{Zero-temperature part}

As explained above, the NLO magnetization of the CSL state is naturally decomposed into two parts. The first of these descends from the difference of the CSL and normal state free energies. At zero temperature, its entire dependence on the magnetic field comes from the charged pion contribution~\eqref{FT0CSLCharged}. Using eq.~\eqref{HurwitzDer} to work out the derivative with respect to $B$, we thus find
\begin{align} \notag
\frac{\Delta M_\text{1,CSL}^{T=0}}{V} = \frac{\sqrt{B} m_\pi}{\sqrt{2}\pi^2 k} \Biggl\{& \int \frac{\dd \lambda }{f_2(\lambda)} \, 
\Biggl[\frac{3}{2}\,\zeta\bigl(-\tfrac{1}{2},\tfrac{1}{2}+\tfrac{m_\pi^2}{2Bk^2}[\lambda-4-k^2] \bigr)\\ \notag
&-\frac{(\lambda-4-k^2)m_\pi^2}{4B k^2} \,\zeta\bigl(\tfrac{1}{2},\tfrac{1}{2}+\tfrac{m_\pi^2}{2Bk^2}[\lambda-4-k^2] \bigr)\Biggr] \left[\chi_{(\lambda_1,\lambda_2)}-\chi_{(\lambda_3,\lambda_4)}\right] \\ \notag
&+ \int_0^\infty \dd P\Biggl[ \frac{2P}{f_2(P^2+\lambda_5)}\biggl\{ \frac{3}{2}\,\zeta\bigl(-\tfrac{1}{2},\tfrac{1}{2}+\tfrac{m_\pi^2}{2Bk^2}[P^2+\lambda_5-4-k^2] \bigr)\\ \notag
&-\frac{(P^2 +\lambda_5-4-k^2)m_\pi^2}{4Bk^2} \,\zeta\bigl(\tfrac{1}{2},\tfrac{1}{2}+\tfrac{m_\pi^2}{2Bk^2}[P^2+\lambda_5-4-k^2] \bigr) \biggr\}\\ \notag
& + \frac{3}{2}\,\zeta\bigl(-\tfrac{1}{2},\tfrac{1}{2}+\tfrac{m_\pi^2}{2Bk^2}[P^2+k^2] \bigr) \\
&- \frac{(P^2+k^2)m_\pi^2}{4Bk^2} \,\zeta\bigl(\tfrac{1}{2},\tfrac{1}{2}+\tfrac{m_\pi^2}{2Bk^2}[P^2+k^2] \bigr) \Biggr]\Biggr\}.\label{DeltaMT0CSL}
\end{align}
Note that this expression diverges when the magnetic field approaches the critical value for BEC of charged pions~\eqref{BBEC} due to the property~\eqref{limZeta}.

On the other hand, the NLO magnetization of the QCD vacuum at zero temperature can be extracted, e.g., from ref.~\cite{Andersen2012b},
\begin{align}\notag
\frac{M_\text{1,vac}^{T=0}}{V} = \frac{m_\pi^2}{16\pi^2}\Biggl\{&2 \,\zeta^{(1,1)}\bigl(-1,\tfrac{1}{2}+\tfrac{m_\pi^2}{2B}\bigr) -\frac{m_\pi^2}{2B}\\
&-\frac{B}{m_\pi^2}\left[8 \,\zeta^{(1,0)}\bigl(-1,\tfrac{1}{2}+\tfrac{m_\pi^2}{2B}\bigr)+\frac{1}{6}\left(2\log\frac{m_\pi^2}{2B}-1\right)\right]\Biggr\}.
\label{MT0vac}
\end{align}
Here the two superscripts of $\zeta$ refer to the order of derivative with respect to the first and second argument of the Hurwitz $\zeta$-function, respectively. 

The complete NLO contribution to the magnetization of the CSL at zero temperature is given by the sum of eqs.~\eqref{DeltaMT0CSL} and~\eqref{MT0vac},
\begin{equation}
\frac{M_\text{1,CSL}^{T=0}}{V} = \frac{\Delta M_\text{1,CSL}^{T=0}}{V} + \frac{ M_\text{1,vac}^{T=0}}{V}.
\label{MT0CSL}
\end{equation}


\subsubsection{Thermal part}

The thermal part of the NLO magnetization can be calculated more straightforwardly than its zero-temperature counterpart. Namely, the contribution of the QCD vacuum to the $B$-dependent part of the thermal free energy~\eqref{FTCSLCharged} is recognized as the second line thereof. Consequently, according to~\eqref{defMag}, the full thermal contribution to the NLO magnetization of the CSL state can be calculated by differentiation of the first line of eq.~\eqref{FTCSLCharged},
\begin{align} \notag
\frac{M_\text{1,CSL}^{T}}{V} = \frac{m_\pi T}{k\pi} \sum_{m=0}^\infty \Biggl\{&\int \frac{\dd \lambda }{\pi f_2(\lambda)} \left[ \log\left[1-e^{-\beta \epsilon_{\text{CSL}}(m,\lambda)}\right] + \frac{(2m+1)B}{2T\epsilon_{\text{CSL}}(m,\lambda)\left(e^{\beta \epsilon_{\text{CSL}}(m,\lambda)}-1\right)}\right]\\
&\times \left[\chi_{(\lambda_1,\lambda_2)}-\chi_{(\lambda_3,\lambda_4)}+\chi_{(\lambda_5,\infty)}\right] \Biggr\}.
\label{MTCSL}
\end{align}
Also this expression diverges in the limit $B\to B_\text{BEC}$.


\section{Numerical results}
\label{sec:results}

\subsection{Phase diagram}\label{sec:PhaseDiag}

With all the analytical results at hand, we now revisit the phase diagram of $\chi$PT at nonzero baryon chemical potential $\mu$, magnetic field $B$ and temperature $T$, published previously in ref.~\cite{Brauner:2021sci}. As pointed out in section~\ref{sec:expF}, we do not have a self-consistently determined free energy as a functional of an arbitrary pion field configuration. As a consequence, we are not able to find the exact ground state, or even its NLO approximation. What we are able to calculate is the NLO correction to the free energy of a field configuration, found by solving the LO equation of motion.

\begin{figure}
\begin{center}
\includegraphics[width=0.8\textwidth]{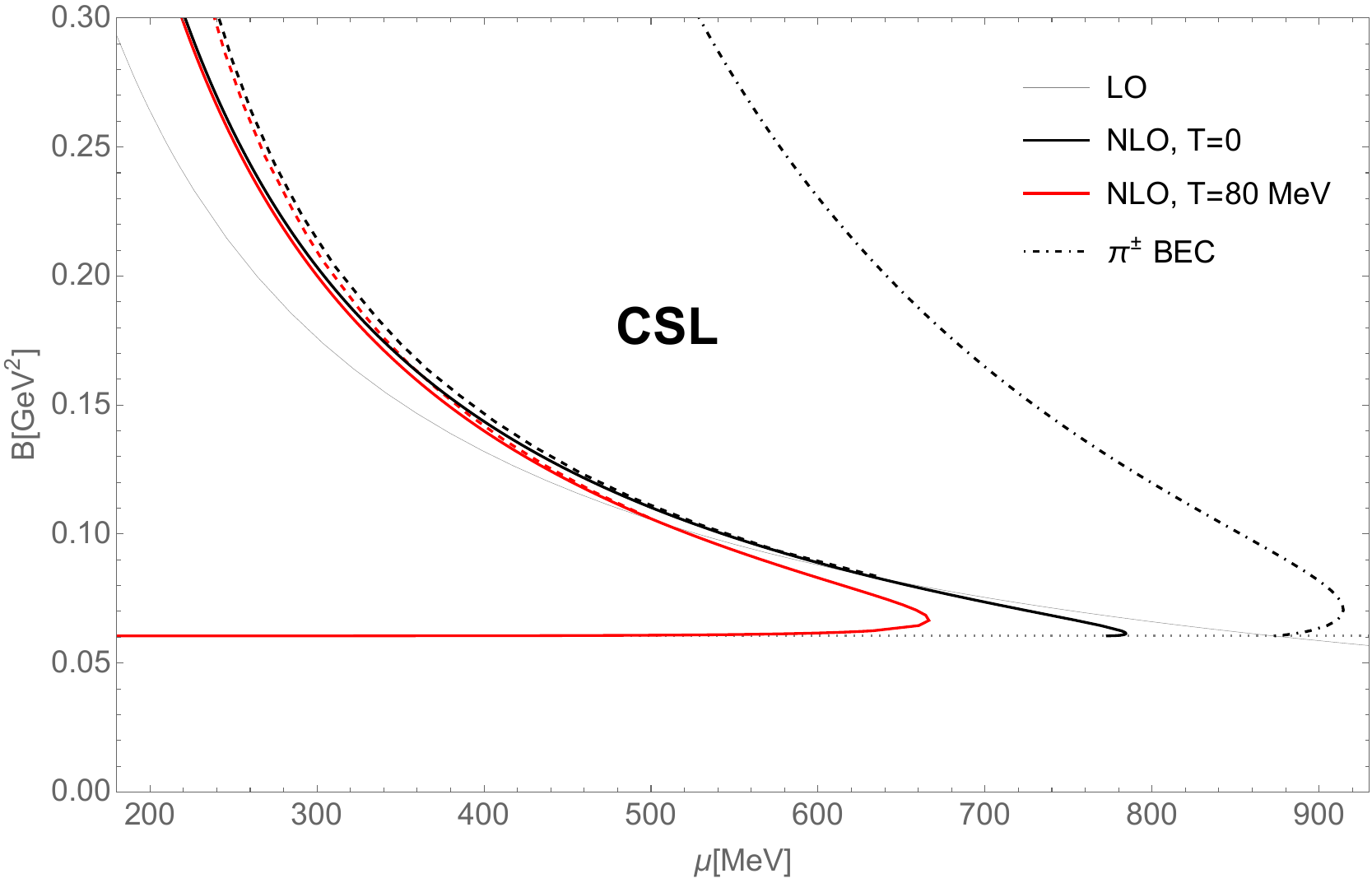}
\caption{Phase diagram of $\chi$PT at nonzero baryon chemical potential $\mu$ and magnetic field $B$ and fixed temperature. The thin gray line corresponds to the LO phase transition between the normal and CSL phases. The solid and dashed lines indicate the same phase transition after having taken into account NLO corrections to the free energy using two different approaches. If one assumes the transition to proceed via the formation of a domain wall, one obtains the solid lines (``domain wall method''). Alternatively, the dashed lines mark the points where the NLO free energy~\eqref{FCSLfull} evaluated on the LO CSL ground state drops to zero (``CSL method''). The black and red lines correspond, respectively, to $T=0$ and $T=80\text{ MeV}$. Furthermore, the dash-dotted line is defined by eq.~\eqref{BBEC} and marks the instability of the charged pion spectrum on the LO CSL background. Finally, the magnetic field $B=3m_\pi^2$ below which the domain wall becomes unstable is represented by the dotted line.}
\label{fig:PhaseDiag}
\end{center}
\end{figure}

With this in mind, we deploy two different strategies to locate the phase transition between the normal and CSL phases, and discuss their mutual consistency. The first strategy was used in ref.~\cite{Brauner:2021sci}, and is based on the assumption that the phase transition proceeds via the formation of a domain wall. The phase boundary is then defined by the condition of vanishing free energy on the domain wall background. Within this ``domain wall method,'' one can explicitly solve for the critical chemical potential at the phase transition as a function of the magnetic field and temperature,
\begin{equation}\label{muCSL}
\mu_\text{CSL}=\frac{16\pi m_\pi f_\pi^2}B+\frac{2\pi}B\frac{\Fa_\text{1,wall}}S,
\end{equation}
where the NLO contribution to the domain wall free energy is given by eq.~\eqref{FwallNLO}. The phase boundary obtained in this way is plotted in figure~\ref{fig:PhaseDiag}, the black and red solid lines corresponding, respectively, to $T=0$ and $T=80\text{ MeV}$. These results show that the CSL state is stabilized by thermal corrections. 

Our second strategy relaxes the assumption that the ground state at the phase transition is the domain wall. What we do instead is to stick to the LO CSL state~\eqref{solCSL}, where the elliptic modulus $k$ is fixed in terms of the chemical potential and magnetic field by the condition~\eqref{condk}. For this state, the free energy up to order $\epsilon^1$ in the chiral counting can be obtained by summing the $\Fa_0$ and $\Fa_1$ pieces as explained in section~\ref{sec:expF}. Consequently, the NLO phase boundary can be approximated by the set of points $(\mu,B)$ at which the complete  free energy~\eqref{FCSLfull} of the LO ground state vanishes: $\Fa_\text{CSL}=0$. The drawback of this ``CSL method'' is that eq.~\eqref{condk} cannot be satisfied for any $B<B_\text{CSL}$. Hence, the part of the parameter space with $B<B_\text{CSL}$ (at given $\mu$) is out of reach for this method, although the ``domain wall method'' suggests that CSL might be favored for $B<B_\text{CSL}$ at sufficiently large chemical potentials.

Numerical results for the phase boundary using the ``CSL method'' are shown in figure~\ref{fig:PhaseDiag} as the black and red dashed lines for, respectively, $T=0$ and $T=80\text{ MeV}$. Also this method confirms that the CSL state is stabilized by thermal fluctuations: the red dashed line always lies below the black one.

Let us now argue that the two methods described above give consistent results. First, note that the dashed and solid lines of the same color (that is, corresponding to the same temperature) meet at points where the domain wall is the LO ground state (the gray line\footnote{Here we choose to evaluate both $B_\text{CSL}$ and $B_\text{BEC}$ using values of the parameters $m_\pi$ and $f_\pi$~\eqref{NLOparameters} fixed by NLO matching. This allows for an easier interpretation of the difference between the displayed LO and NLO phase transitions. This is in contrast to ref.~\cite{Brauner2017a}, where the physical values of pion mass and decay constant were used to plot the phase diagram.} in figure~\ref{fig:PhaseDiag}). Specifically, the zero-temperature phase boundaries (black lines) meet at $(\mu,B)=(631\text{ MeV},0.0837\text{ GeV}^2)$, whereas the $T=80\text{ MeV}$ phase boundaries (red lines) meet at $(\mu,B)=(506\text{ MeV},0.104\text{ GeV}^2)$. Second, for smaller chemical potentials, each of the dashed lines lies above the corresponding solid line. In other words, along the phase boundary obtained with the ``CSL method'' (dashed lines), the domain wall state has lower NLO free energy than the LO CSL ground state. This indicates that the phase transition may indeed proceed via the formation of a domain wall. 

Finally, the dash-dotted line in figure~\ref{fig:PhaseDiag} corresponds to the magnetic field $B_\text{BEC}$~\eqref{BBEC}; this marks the instability of the charged pion spectrum on the CSL background minimizing the LO free energy. The corresponding instability for the domain wall background appears at $B=3m_\pi^2$ and is displayed in  figure~\ref{fig:PhaseDiag} as the dotted line. The fact that the thermal contribution to the NLO free energy of the domain wall diverges for $B\to 3m_\pi^2$ explains why the ``domain-wall-method'' phase boundary at nonzero temperature  appears to move towards vanishing baryon chemical potential in this limit. We stress, however, that this result should be taken with a grain of salt. Namely, such a large deviation of the NLO phase boundary from the LO one hints at a large one-loop correction to the free energy, which may merely signal a breakdown of the derivative expansion of $\chi$PT.

We would like to stress that this work focuses solely on the phase transition between the vacuum and CSL phases appearing at $B\gtrsim 3m_\pi^2$. The points of instability of the charged pion spectrum are marked in figure~\ref{fig:PhaseDiag} only for illustration; the exact location of the charged pion BEC phase transition and the form of the new ground state featuring a condensate of charged pions are beyond the scope of our work. We come back to this point in the discussion (see section~\ref{sec:summary}).


\subsubsection{Chiral limit}

Let us briefly comment on the limit of vanishing pion mass, which is for simplicity often assumed in literature on inhomogeneous chiral phases of quark matter. In the chiral limit, the LO CSL ground state simplifies to
\begin{equation}\label{GroundChir}
\phi_0(z)=\frac{\mu B z}{4\pi^2 f_\pi^2}.
\end{equation}
The energy density of this state relative to the QCD vacuum is
\begin{equation}\label{ECSLchir}
\frac{\Fa_\text{0,CSL}}{V}\biggr|_{m_\pi=0} =- \frac{\mu^2 B^2}{32 \pi^4 f_\pi^2}.
\end{equation}
The CSL state thus becomes energetically favored over the normal state for arbitrarily small $\mu$ and $B$. We have checked that this remains true when the NLO contribution is added to the free energy (see appendix~\ref{app:chiral} for explicit results). As a consequence, the phase diagram becomes significantly simpler in the chiral limit. The only nontrivial phase boundary that persists is the one corresponding to BEC of charged pions, given in the chiral limit by eq.~\eqref{BBECchiral}.


\subsection{Magnetization}

Here we illustrate numerically the results obtained in sections~\ref{sec:MagWall} and~\ref{sec:MagCSL}.

\subsubsection{Domain wall}

\begin{figure}
\begin{center}
\includegraphics[width=0.8\textwidth]{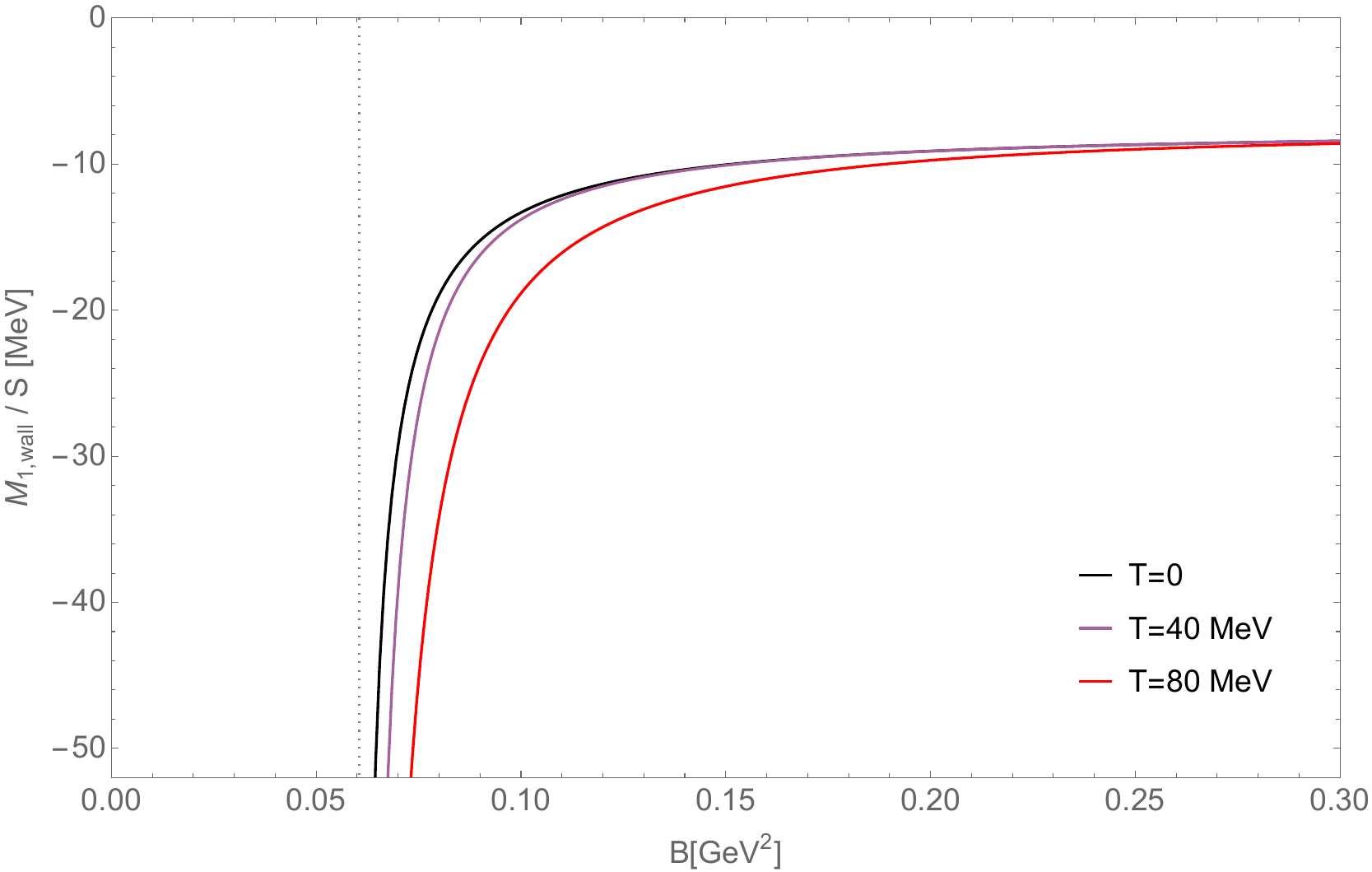}
\caption{NLO contribution to the magnetization of the domain wall per unit surface given by eqs.~\eqref{magWallT0} and~\eqref{magWallT}. The black, purple and red line corresponds respectively to $T=0$, $T=40\text{ MeV}$ and $T=80\text{ MeV}$. Note that in order to obtain the complete magnetization of the domain wall, the LO contribution $\mu/2\pi$ has to be added.}
\label{fig:magDW}
\end{center}
\end{figure}

The LO magnetization per unit surface of the domain wall equals $\mu/(2\pi)$, independently of the magnetic field. On the other hand, the NLO contributions to the magnetization~\eqref{magWallT0} and~\eqref{magWallT} are independent of $\mu$. We therefore choose to plot solely the NLO contribution to the magnetization as a function of $B$, see figure~\ref{fig:magDW}.

\subsubsection{Chiral soliton lattice}

\begin{figure}
\begin{center}
\includegraphics[width=0.8\textwidth]{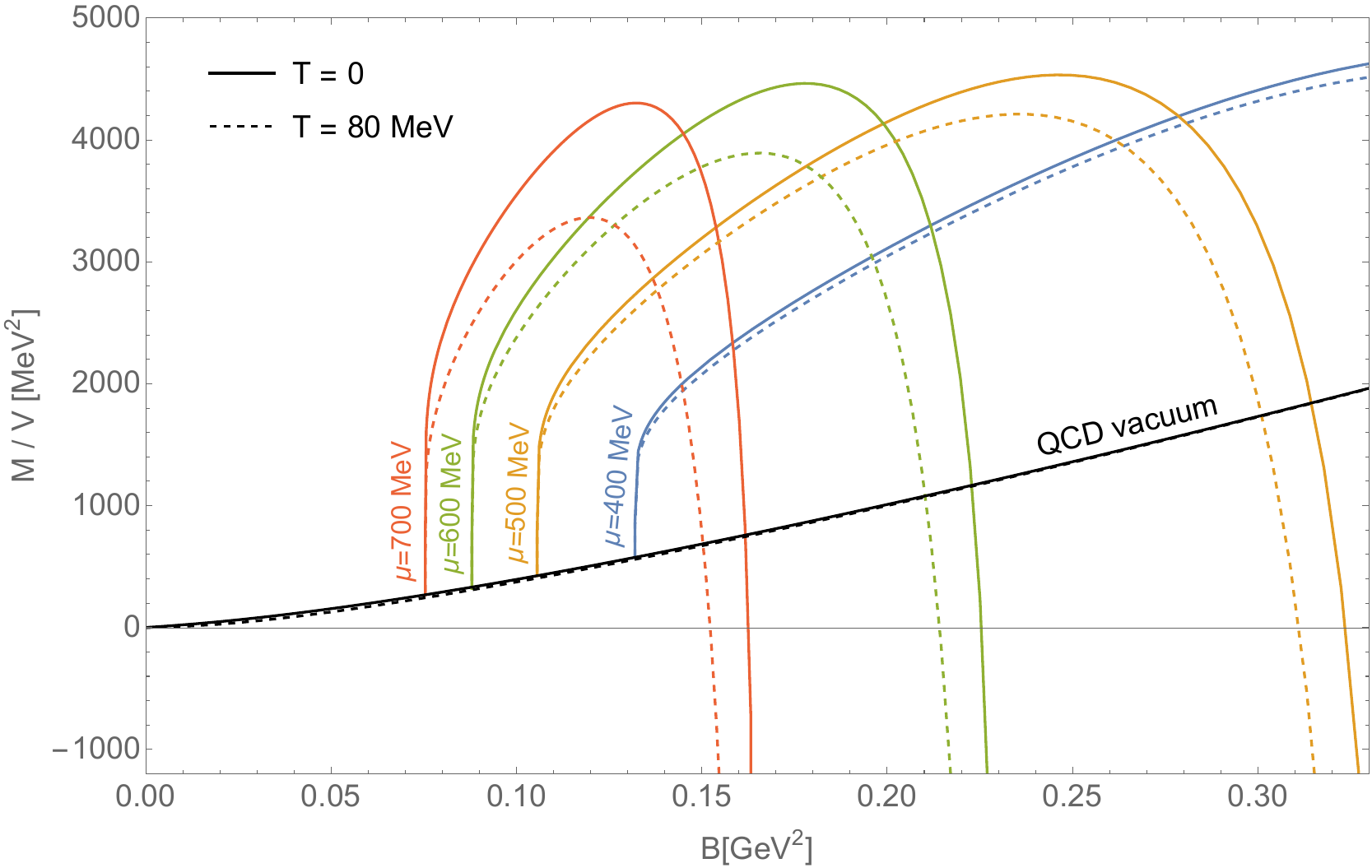}
\caption{The complete (LO plus NLO) magnetization of the CSL per unit volume as given by the sum of eqs.~\eqref{MCSLLO}, \eqref{MT0CSL} and~\eqref{MTCSL}. The solid lines show numerical values for $T=0$; the blue, yellow, green and red color corresponds respectively to $\mu$ equal to $400$, $500$, $600$ and $700 \text{ MeV}$. The dashed lines of the same colors correspond to the temperature of $80\text{ MeV}$ and the same $\mu$. For given $\mu$ and magnetic field $B$, the elliptic modulus $k$ is fixed by the condition~\eqref{condk} on the LO CSL ground state. For comparison, the black solid and dashed lines show the magnetization of the QCD vacuum~\cite{Andersen2012b} for $T=0$ and $T=80\text{ MeV}$, respectively.}
\label{fig:magCSL}
\end{center}
\end{figure}

The LO magnetization per unit cell of the lattice and unit transverse area equals $\mu/(2\pi)$ analogously to the domain wall. However, the spatially averaged magnetization per unit \emph{volume} depends implicitly on the magnetic field through the value of the elliptic modulus $k$, fixed by eq.~\eqref{condk}. For the same reason, the NLO contributions to magnetization~\eqref{MT0CSL} and~\eqref{MTCSL} depend implicitly on the baryon chemical potential. In figure~\ref{fig:magCSL}, we therefore show the complete magnetization of the CSL state,
\begin{equation}\label{MCSLfull}
\frac{M_\text{CSL}}{V}=\frac{M_\text{0,CSL}}{V}+\frac{M_\text{1,CSL}^{T=0}}{V}+\frac{M_\text{1,CSL}^{T}}{V},
\end{equation}
as a function of the magnetic field for several values of the baryon chemical potential. The solid and dashed lines correspond to $T=0$ and $T=80\text{ MeV}$, respectively.

For comparison, we also show in figure~\ref{fig:magCSL} the magnetization density of the QCD vacuum (black solid and dashed lines). Note that as the magnetic field approaches the critical value $B_\text{CSL}$ from above, the magnetization of the CSL converges, not surprisingly, towards that of the vacuum. To visualize this feature, we display the magnetization for the values of $B$ all the way to the LO phase boundary although, strictly speaking, the CSL state is not energetically favorable at NLO for the lowest magnetic fields in case of $\mu=400$, $500$ and $600\text{ MeV}$. For example, in case of $\mu=400\text{ MeV}$, the LO phase boundary appears for $B=0.132\text{ GeV}^2$ whereas the NLO phase boundary for $T=0$ and $T=80\text{ MeV}$ appears respectively for $B=0.144\text{ GeV}^2$ and $B=0.140\text{ GeV}^2$ within the ``domain wall method''.

\section{Summary and discussion}
\label{sec:summary}

Sufficiently strong magnetic fields and large baryon chemical potentials trigger the formation of a periodic condensate of neutral pions out of the QCD vacuum. In ref.~\cite{Brauner2017a}, the domain in the QCD phase diagram, occupied by this novel CSL state of matter, was determined at the LO of the derivative expansion of $\chi$PT, which restricts the validity of the analysis to zero temperature. In our preceding paper~\cite{Brauner:2021sci} and here, we extended the analysis to NLO by including the one-loop corrections induced by fluctuations above the CSL state. The main conclusion is that the CSL phase is stabilized by thermal fluctuations.

We used two different strategies to locate the phase transition between the normal and CSL phases. In the first approach, we assumed that the phase transition proceeds via domain wall formation, and defined the phase boundary by the condition $\Fa_\text{wall}=0$. In the second approach, we evaluated the NLO free energy on the LO ground state at given magnetic field and chemical potential, and then imposed the condition $\Fa_\text{CSL}=0$. Within the part of the parameter space where both these methods are applicable, they agree up to an error expected in our perturbative expansion of the free energy~\eqref{FNLO}. Namely, the difference between the two predictions for the phase boundary is much smaller than the difference between the LO and NLO phase boundaries; see figure~\ref{fig:PhaseDiag}. Moreover, the position of the phase boundary obtained using the ``CSL method'' suggests that the phase transition may indeed proceed via the formation of a domain wall.

As observed already in ref.~\cite{Brauner2017a}, the lowest-lying charged pion excitation of the CSL state becomes massless at the critical magnetic field $B_{\text{BEC}}$~\eqref{BBEC}, which indicates an instability of CSL under BEC of charged pions. The question what the new, even more energetically favored ground state at $B>B_\text{BEC}$ is, was addressed in ref.~\cite{Evans:2022hwr}. It was shown that, in analogy with type-II superconductors, the ground state configuration features a periodic array of magnetic flux tubes and a periodic condensate of charged pions.

The analysis of ref.~\cite{Evans:2022hwr} is, however, restricted to the limit of vanishing pion mass. While its results therefore cannot be directly applied to the real world where pions are massive, one can make at least an educated guess based on the hierarchy of the different scales at play. For $B\gg m_\pi$, the chiral limit gives a reasonably accurate description of the interplay between the CSL and charged pion BEC. Hence, we expect the Abrikosov vortex lattice carrying a charged pion BEC to appear at $B\gtrsim B_\text{BEC}$. On the other hand, the features of the phase diagram in figure~\ref{fig:PhaseDiag} arising from the instability of the domain wall for $B<3m_\pi^2$ are clearly not accessible by a chiral limit analysis. We do know that at very low $B$, the QCD vacuum will prevail. It would therefore be very interesting to explore whether and where a transition between this normal phase and the BEC of charged pions appears.

The CSL is not the only inhomogeneous phase proposed to appear in the phase diagram of QCD. While all our results have been obtained using model-independent effective field theory and thus constitute genuine predictions of QCD within the limitations of the derivative expansion of $\chi$PT, it is nevertheless instructive to compare these predictions with other works, typically based on simplified models of QCD. In refs.~\cite{Tatsumi:2014wka,Abuki:2018iqp}, the effect of a magnetic field on the ground state of QCD in presence of the chiral anomaly was studied using a Ginzburg-Landau-type approach. This generally leads to the prediction of an inhomogeneous phase that appears in the literature under different names such as ``chiral spiral'' or ``(dual) chiral density wave,'' and coincides with the CSL in the chiral limit. The same problem was addressed in refs.~\cite{Nishiyama:2015fba,Ferrer2015a} using the Nambu-Jona-Lasinio model, yet only in the chiral limit. Our results generally agree with such model studies in the part of the phase diagram, accessible to $\chi$PT in the derivative expansion, that is for chemical potentials well below the onset of nuclear matter.

Finally, note that the original prediction of the CSL phase in QCD under strong magnetic fields~\cite{Brauner2017a} was subsequently extended to related setups where either the external conditions or the dynamical theory itself may differ. This applies in particular to dense QCD matter under uniform rotation in presence of a baryon or isospin chemical potential or both~\cite{Huang2018a,Nishimura:2020odq,Eto:2021gyy}, QCD in external magnetic field and with nonzero isospin chemical potential~\cite{Gronli:2022cri,Adhikari:2022cks}, and a class of QCD-like theories with nonzero magnetic field and baryon chemical potential~\cite{Brauner2019c,Brauner2019b}. We expect the computational techniques developed here to be useful also in these other contexts, should one be interested in the effects of quantum or thermal fluctuations therein.


\acknowledgments

We are indebted to Naoki Yamamoto for a collaboration preceding this project. We are also grateful to Geraint Evans, Andreas Schmitt, Gilberto Colangelo and Martina Cottini for helpful discussions. This work was supported in part by the University of Stavanger and the University Fund under the grant no.~PR-10614. HK is further supported in part by the Swiss National Science Foundation (SNSF) under grant no.~200020B-188712.


\appendix

\section{Spectrum of effective one-dimensional Hamiltonians} \label{app:spectra}

\subsection{P\"oschl-Teller Hamiltonian} \label{subsec:PTHamiltonian}

The Hamiltonians~\eqref{Hamwall} and~\eqref{Hamwallcharged} are both special cases of the P\"oschl-Teller Hamiltonian,
\begin{equation}\label{HamPoschl}
\Ha_n\equiv-\de_{\bar{z}}^2-\frac{n(n+1)}{\cosh^2\bar{z}}.
\end{equation}
Let us briefly recall how to find the eigenvalues and eigenstates of this class of Hamiltonians for positive integer $n$. Introducing a set of annihilation and creation operators,
\begin{equation}
a_n\equiv\frac\dd{\dd \bar{z}}+n\tanh \bar{z},\quad
\he a_n\equiv-\frac\dd{\dd \bar{z}}+n\tanh \bar{z},
\end{equation}
the Hamiltonian $\Ha_n$ can be expressed as
\begin{equation}
\Ha_n=\he a_na_n-n^2=a_{n+1}\he a_{n+1}-(n+1)^2.
\end{equation}
This implies a pair of important intertwining relations,
\begin{equation}
a_n\Ha_n=\Ha_{n-1}a_n,\qquad
\he a_n\Ha_{n-1}=\Ha_n\he a_n.
\end{equation}
Using these relations, one can prove the following basic facts about the spectrum of the P\"oschl-Teller Hamiltonians:
\begin{itemize}
\itemsep=0pt
\item The Hamiltonian $\Ha_n$ has $n$ bound states $\ket{n,k}$, labeled by an integer $k$, $1\leq k\leq n$. The corresponding eigenvalue is $\lambda_{n,k}=-(n-k+1)^2$.
\item The continuous spectrum of $\Ha_n$ consists of doubly degenerate energy levels covering the open set $(0,\infty)$.
\item The eigenstates in the continuous spectrum are obtained by applying a chain of creation operators on a plane wave,
\begin{equation}
\ket{n,P}\propto\he a_n\dotsb\he a_1\ket{0,P},
\label{eigenstates}
\end{equation}
where $\braket {\bar{z}}{0,P}\equiv e^{\imag P\bar{z}}$. The corresponding eigenvalue of the Hamiltonian is
\begin{equation}
\lambda_{n,P}=P^2.
\label{LambdaP2}
\end{equation}
\item The eigenstates $\ket{n,P}$ have the asymptotic behavior of reflectionless scattering states with momentum $P$. The corresponding phase shift $\delta_P$ is given by
\begin{equation}
e^{\imag\delta_P}=\frac{\imag P-1}{\imag P+1}\dotsb\frac{\imag P-n}{\imag P+n}.
\label{deltaP}
\end{equation}
\end{itemize}
As we will now show, the information about the phase shift $\delta_P$ is sufficient to convert the sum over eigenvalues of the P\"oschl-Teller Hamiltonian into an integral over the continuous momentum variable $P$.

Let us for simplicity focus on the simplest case of $n=1$, relevant for neutral pion fluctuations of the domain wall; the $n=2$ case relevant for charged pion fluctuations can be dealt with in exactly the same manner. In the case of $n=1$, the phase shift $\delta_P$ is a monotonously decreasing function of $P$ such that
\begin{equation}
\lim_{P\to0+}\delta_P=\pi,\qquad
\lim_{P\to+\infty}\delta_P=0.
\end{equation}
This is consistent via Levinson's theorem with the fact that the Hamiltonian $\Ha_1$ has a single bound state.

To be able to count the states in the continuous spectrum, we temporarily enclose our system in a finite box, $-L/2\leq\bar z\leq+L/2$, and impose the Dirichlet boundary condition. In order to obtain a real eigenstate of $\Ha_1$ satisfying this boundary condition, we have to take a linear combination of the two states $\ket{1,\pm P}$, where $P$ is from now on implicitly assumed to be positive. Assuming that $L$ is sufficiently large so that the exponentially decaying tails of the eigenstates~\eqref{eigenstates} can be neglected, the boundary condition at $\bar z=-L/2$ can be satisfied by taking
\begin{equation}
\begin{aligned}
\psi_P(\bar{z})&=\sin[P(\bar{z}+L/2)]&\text{ for }&\bar{z}\to-\infty,\\
\psi_P(\bar{z})&=\sin[P(\bar{z}+L/2)+\delta_P]\quad&\text{ for }&\bar{z}\to+\infty.
\end{aligned}
\end{equation}
On the second line, we used the fact that $\delta_P$ is an odd function of $P$ up to a multiple of $2\pi$. The boundary condition at $\bar z=+L/2$ then isolates a discrete set of standing-wave eigenstates, labeled by integer $N$ such that
\begin{equation}
PL+\delta_P=N\pi.
\label{standing1}
\end{equation}
For sufficiently large $L$, the left-hand side is a monotonously increasing function of $P$ that approaches $\pi$ as $P\to0+$. Hence the discrete set of eigenstates satisfying our boundary condition is labeled by integers $N\geq2$.\footnote{Note that $N=1$ corresponds to the ``half-bound state'' with $P=0$. Being nondegenerate, this cannot satisfy the boundary condition exactly.} The condition~\eqref{standing1} is to be compared to the corresponding condition for the ``free-particle'' Hamiltonian $\Ha_0$,
\begin{equation}
PL=N\pi,
\label{standing2}
\end{equation}
which gives a normalizable standing-wave eigenstate of $\Ha_0$ for any integer $N\geq1$.

Suppose now that we want to evaluate the sum of some given function $f(\lambda)$ over all eigenvalues $\lambda$ of the Hamiltonian $\Ha_1$. In order to ameliorate the expected divergence of the sum, we will subtract the corresponding sum over eigenvalues of the ``free-particle'' Hamiltonian $\Ha_0$. Given the relation~\eqref{LambdaP2}, the result, which we will for the sake of brevity call $\tr f$, then reads
\begin{equation}
\tr f=f(-1)-f(0)+\sum_{N=1}^\infty[f(P_N^2)-f(\tilde P_N^2)].
\end{equation}
Here $P_N$ is the solution of eq.~\eqref{standing1}, and $\tilde P_N$ the solution of the corresponding condition~\eqref{standing2} for the free particle. Since only $N\geq2$ give a stationary state satisfying the Dirichlet boundary condition, the $N=1$ contribution, for which $P_1=0$, has to be explicitly subtracted. Finally, the first term, $f(-1)$, is the contribution of the bound state.

As the next step, we will convert the sum into an integral over the dimensionless momentum $P$ by taking the limit $L\to\infty$. Note that upon taking the difference of eqs.~\eqref{standing1} and~\eqref{standing2}, we get
\begin{equation}
(P_N-\tilde P_N)L=-\delta_{P_N}.
\end{equation}
For $L\to\infty$, $P_N-\tilde P_N$ is infinitesimally small so that we can approximate
\begin{equation}
f(P_N^2)-f(\tilde P_N^2) \approx(P_N^2-\tilde P_N^2)f'(\tilde P_N^2)\approx 2\tilde P_N(P_N-\tilde P_N)f'(\tilde P_N^2) =-\frac{2\tilde P_N\delta_{P_N}}{L}f'(\tilde P_N^2).
\end{equation}
The sum over $N$ corresponds to the spacing of momentum $\Delta\tilde P=\pi/L$, which gives the correct measure for the integral over $P$, the result being
\begin{equation}
\tr f=f(-1)-f(0)-\frac2\pi\int_0^\infty P\delta_Pf'(P^2)\,\dd P.
\end{equation}
This expression can be further simplified by integration by parts. The boundary contribution at $P=0$ exactly cancels the obnoxious $-f(0)$ term before the integral. Upon further using the fact that $\dd\delta_P/\dd P=-2/(1+P^2)$, we arrive at our final result for the $n=1$ P\"oschl-Teller Hamiltonian,
\begin{equation}
\tr f= f(-1)-\frac1\pi\lim_{P\to\infty}[\delta_Pf(P^2)]-\frac2\pi\int_0^\infty\frac{f(P^2)}{1+P^2}\,\dd P,\qquad(n=1).
\label{trf}
\end{equation}
Given the asymptotic behavior of the phase shift at $P\to\infty$, $\delta_P=2/P+\dotsb$, the remaining surface term vanishes for functions such that $\lim_{P\to\infty}[f(P^2)/P]=0$.

The evaluation of the sum over eigenvalues in the $n=2$ case follows exactly the same steps, the result being
\begin{align}
\label{trfcharged}
\tr f={}&f(-4)+f(-1)-\frac1\pi\lim_{P\to\infty}[\delta_Pf(P^2)]\\
\notag
&
-\frac1\pi\int_0^\infty\left(\frac2{1+P^2}+\frac4{4+P^2}\right)f(P^2)\,\dd P,\qquad(n=2).
\end{align}


\subsection{Lam\'e Hamiltonian}

\begin{figure}
\begin{center}
\includegraphics[width=0.8\textwidth]{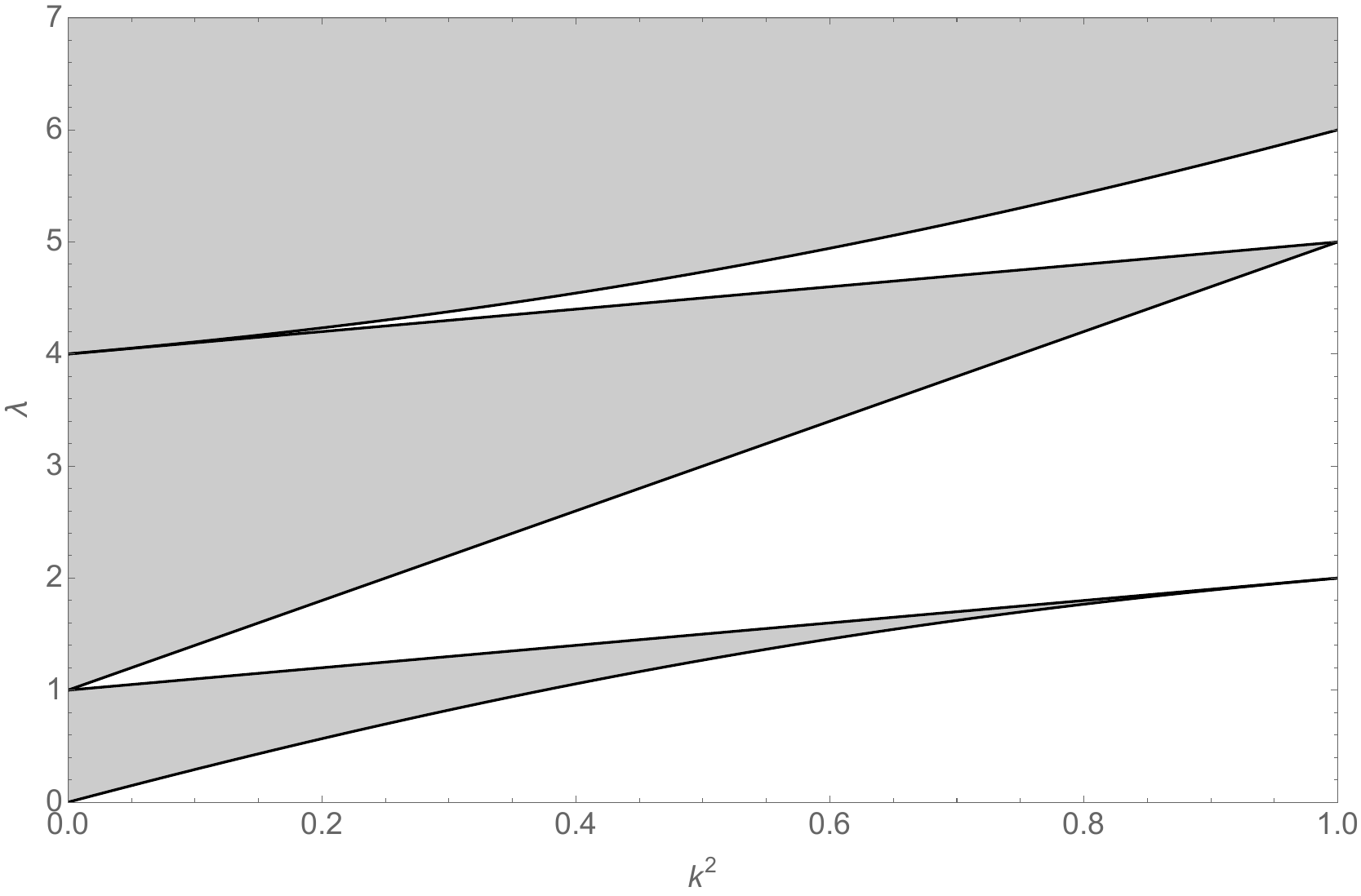}
\caption{Dependence of the band spectrum of the $n=2$ Lam\'e Hamiltonian~\eqref{HamLame} on the elliptic modulus $k$. For $k\to 1$, the two lowest-lying bands collapse to two discrete bound states. For $k\to 0$, the spectrum of the free Hamiltonian is recovered. The exact expressions for the five edges of the three bands are given in eq.~\eqref{lambdasCharged}.}
\label{fig:specLame}
\end{center}
\end{figure}

The Hamiltonians~\eqref{HamCSL} and~\eqref{HamCSLcharged} are both special cases of the Lam\'e Hamiltonian
\begin{equation}\label{HamLame}
\Ha_n \equiv -\de_{\bar{z}}^2+ n(n+1) k^2\sn^2(\bar{z},k). 
\end{equation}
In general, the spectrum of these Hamiltonians is known to consist of $n+1$ energy bands. In the limit $k\to1$, $\sn(\bar z,k)\to\tanh\bar z$. As a consequence, the Lam\'e Hamiltonian~\eqref{HamLame} reduces to the P\"oschl-Teller Hamiltonian~\eqref{HamPoschl} up to a constant shift, $n(n+1)$. Accordingly, the lowest $n$ bands of the Lam\'e Hamiltonian collapse to $n$ discrete energy levels, corresponding to the bound states of the P\"oschl-Teller Hamiltonian. In the opposite limit of $k\to0$, the Lam\'e Hamiltonian tends to the ``free-particle'' Hamiltonian $-\de_{\bar z}^2$, hence the gaps between the different bands close. To visualize how the spectrum interpolates between the two limits, that is for $0<k<1$, we display in figure~\ref{fig:specLame} the band structure in the special case of $n=2$, as a function of $k$. The expressions for the edges of the energy bands, given in eqs.~\eqref{bandsneutral} and~\eqref{lambdasCharged}, were extracted from refs.~\cite{Li2000a,Maier2008a}.

The Lam\'e Hamiltonian has no bound states. Accordingly, all the energy levels can be parameterized by a dimensionless momentum variable $P$. Owing to the periodicity of the potential in the Hamiltonian, this is identified with the crystal (Bloch) momentum of the eigenstate. An explicit expression for the eigenvalue $\lambda$ as a function of $P$ is not known. For the purposes of this paper, it is however sufficient to know the group velocity $\dd\lambda/\dd P$, given in eqs.~\eqref{DoSneutral} and~\eqref{DoScharged}~\cite{Li2000a,Maier2008a}. 

For later convenience, let us add that in the special case of the $n=1$ Lam\'e Hamiltonian, the spectrum can also be specified parametrically in terms of a complex parameter $\alpha$~\cite{Li2000a},
\begin{equation}\label{SpecLamePar}
\lambda(\alpha)=\dn^2\alpha+k^2,\qquad
P(\alpha)=-\imag Z(\alpha) + \frac{\pi}{2K(k)}.
\end{equation}
It can be shown that the function $P(\alpha)$ defined by eq.~\eqref{SpecLamePar} is real either for purely imaginary $\alpha=\imag\eta$ or for $\alpha=K(k)+\imag\eta$. The two options with $\eta\in (0,K(k'))$ span, respectively, the states in the upper and lower bands of the spectrum of $\Ha_1$.


\section{Spectrum of excitations above the CSL background}
\label{app:phonon}

With the insight in the spectrum of the Lam\'e Hamiltonian gained in appendix~\ref{app:spectra}, it is possible to be more explicit about the dispersion relations of the fluctuations of the CSL. It follows from eqs.~\eqref{F1CSL} and~\eqref{F2CSL} that the energy (frequency) of neutral and charged-pion fluctuations is given, respectively, by
\begin{equation}
\begin{split}
\omega_{\pi^0}^2&=\vek p_\perp^2+m_\pi^2\left[\frac{\lambda(P)}{k^2}-1\right],\\
\omega_{\pi^\pm}^2&=(2m+1)B +\frac{m_\pi^2}{k^2}[\lambda(P)-k^2-4].
\end{split}
\end{equation}
The dimensionless momentum $P$ is related to the physical momentum $p_z$ along the magnetic field, $P\equiv kp_z/m_\pi$. The dispersion relations will thus be completely fixed once we know the functions $\lambda(P)$ for the $n=1$ (neutral pions) and $n=2$ (charged pions) Lam\'e Hamiltonians.

As already stressed, a closed expression for $\lambda(P)$ is not known. However, it is possible to determine implicitly $P$ as a function of $\lambda$ by integrating the inverse of the group velocities~\eqref{DoSneutral} and~\eqref{DoScharged}. Below, we demonstrate how to do so explicitly in the case of $n=1$, that is neutral pions. To that end, we use the various representations of elliptic integrals from ref.~\cite{Abramowitz1972a}. The dispersion relations presented below generalize the result of ref.~\cite{Brauner2017a}, where the phase velocity of CSL phonons, that is neutral pion excitations near the bottom of the lower, gapless energy band, was calculated.


\subsection{Neutral pion excitations: lower band}

For the lower, gapless energy band, we find
\begin{equation}
\begin{split}
P&=E(\arcsin\tau,k')+\biggl[\frac{E(k)}{K(k)}-1\biggr]F(\arcsin\tau,k'),\\
\tau&\equiv\frac{\sqrt{\lambda-k^2}}{k'}=\frac k{k'}\frac{\sqrt{\omega^2-\vek p_\perp^2}}{m_\pi},
\end{split}
\end{equation}
where $k' \equiv \sqrt{1-k^2}$ is the complementary elliptic modulus. This gives the dispersion relation parametrically as $\omega=\omega(\vek p_\perp,\tau)$ and $P=P(\tau)$. Expanding the function $P(\tau)$ to linear order in $\tau$ recovers the phonon phase velocity derived in ref.~\cite{Brauner2017a}. The upper edge of the band corresponds to $\lambda=\tau=1$, whence we find the energy
\begin{equation}\label{omEdgeM}
\omega_\text{edge}^-=\sqrt{\vek p_\perp^2+\frac{k'^2}{k^2}m_\pi^2}.
\end{equation}
The corresponding momentum is 
\begin{equation}\label{Pedge}
P_\text{edge}=\frac\pi{2K(k)},\quad\text{or}\quad
p_{z,\text{edge}}=\frac\pi L,
\end{equation}
where $L$ is the lattice spacing of the CSL state~\eqref{CSLperiod}.


\subsection{Neutral pion excitations: upper band}

For the upper energy band, we find likewise
\begin{equation}
\begin{split}
P={}&\frac\pi{2K(k)}+\left[1-\frac{E(k)}{K(k)}\right]F(\arctan\tau,k')-E(\arctan\tau,k')+\tau\sqrt{\frac{1+k^2\tau^2}{1+\tau^2}},\\
\tau\equiv{}&\frac{\sqrt{\lambda-1-k^2}}k=\frac1{m_\pi}\sqrt{\omega^2-\vek p_\perp^2-\frac{m_\pi^2}{k^2}}.
\end{split}
\end{equation}
The lower edge of the band corresponds to $\lambda=1+k^2$ or $\tau=0$, and has the energy
\begin{equation}\label{omEdgeP}
\omega_\text{edge}^+=\sqrt{\vek p_\perp^2+\frac{m_\pi^2}{k^2}}.
\end{equation}
The energy gap at vanishing transverse momentum is therefore
\begin{equation}
\omega_\text{edge}^+-\omega_\text{edge}^-\Bigr|_{\vek p_\perp=\vek0}=\frac{1-k'}km_\pi=\frac k{1+k'}m_\pi.
\end{equation}
It equals $m_\pi$ for the domain wall solution where $k=1$, and then rapidly drops until it eventually disappears in the limit $k\to0$.

\begin{figure}
\begin{center}
\includegraphics[width=0.8\textwidth]{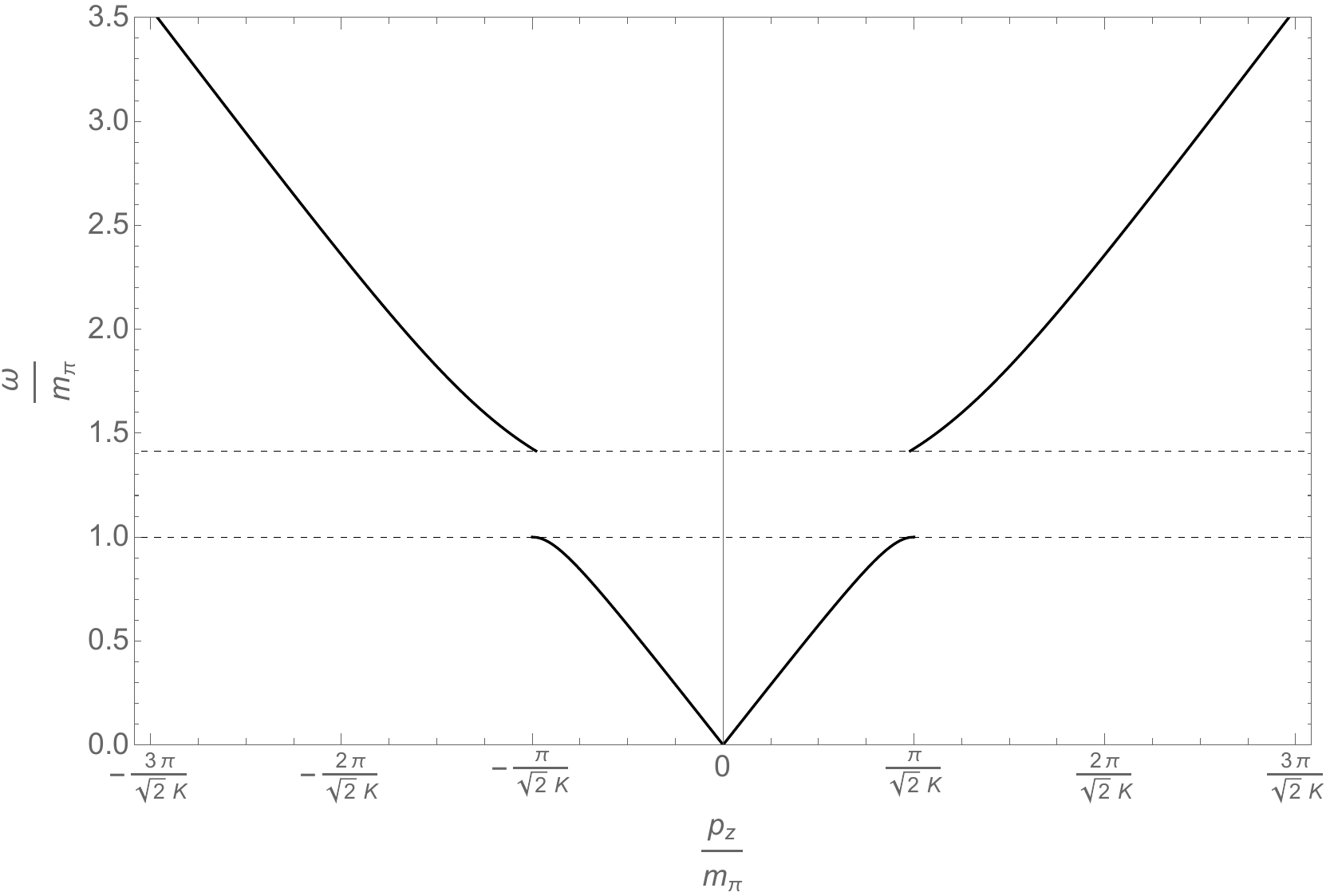}
\caption{Dispersion relation of neutral pion excitations on the CSL background with $k^2=1/2$ and $\vek p_\perp=\vek0$. According to eq.~\eqref{Pedge}, the edges of the first Brillouin zone are at $p_z=\pm \pi m_\pi/(\sqrt{2}K)$ where $K$ is a shorthand for $K(1/\sqrt2)$. The energy gap appears between $\omega = m_\pi$ and $\sqrt{2} m_\pi$ according to eqs.~\eqref{omEdgeM} and~\eqref{omEdgeP}.}
\label{fig:phonon}
\end{center}
\end{figure}

For illustration, we plot the dispersion relation in both bands for $k^2=1/2$ in figure~\ref{fig:phonon}. 

\section{Alternative evaluation of neutral pion free energy at NLO}
\label{app:Gelfand}

The Gelfand-Yaglom theorem is a practically useful tool to evaluate functional determinants of one-dimensional differential operators~\cite{Dunne2008a}. It can help us to efficiently evaluate the sum over $\lambda$ in eq.~\eqref{F1CSL} thanks to the fact that the argument of the logarithm therein is linear in $\lambda$. It is important that this is done before the Matsubara sum or the integral over $\vek p_\perp$ is carried out. Throughout this appendix, we heavily rely on the properties of the Jacobi elliptic functions and further identities that can be found in ref.~\cite{Abramowitz1972a}.

The sum over $\lambda$ in eq.~\eqref{F1CSL} is equivalent to a sum over the eigenvalues of the operator
\begin{equation}
\omega_n^2+\vek p_\perp^2-m_\pi^2+\frac{m_\pi^2}{k^2}\Ha^{(\pi^0)}_\text{CSL}.
\label{Hamaux}
\end{equation}
Finding the latter is in turn equivalent to solving the eigenvalue problem
\begin{equation}
\Ha^{(\pi^0)}_\text{CSL}\psi=(2k^2-\Omega_0^2)\psi,
\label{Hamaux2}
\end{equation}
where $\Omega_0$ is defined by eq.~\eqref{Omega0def}. This is a special case of the Lam\'e equation. According to ref.~\cite{Li2000a}, the two linearly independent solutions of this equation are
\begin{equation}
\psi_\pm(\bar z)\propto\frac{H(\bar z\pm\xi,k)}{\Theta(\bar z,k)}e^{\mp\bar z Z(\xi,k)},
\label{psi}
\end{equation}
where the constant $\xi$ is given implicitly as the solution to the condition
\begin{equation}\label{condAux}
2k^2-\Omega_0^2=k^2+\dn^2(\xi,k).
\end{equation}
This is equivalent to
\begin{equation}
\dn^2(\xi,k)=-\frac{k^2}{m_\pi^2}(\omega_n^2+\vek p_\perp^2).
\end{equation}
It follows that $\dn(\xi,k)$ is necessarily imaginary. This constrains possible values of $\xi$ to
\begin{equation}
\xi=u+\imag(2n+1)K(k'),
\end{equation}
where $n$ is an integer that can without loss of generality be set to zero. This allows one to trade the complex parameter $\xi$ for the real parameter $u$ so that $\dn^2(\xi,k)=-\cs^2(u,k)$, which reproduces the condition~\eqref{csak} for $u$.

Now that the imaginary part of $\xi$ is fixed, the solutions~\eqref{psi} can be cast in a manifestly real form
\begin{equation}
\psi_\pm(\bar z)=\frac{\Theta(\bar z\pm u,k)}{\Theta(\bar z,k)}e^{\mp\bar z\Xi(u,k)},
\end{equation}
where we used the shorthand notation $\Xi(u,k)\equiv Z(u,k)+\cs(u,k)\dn(u,k)$ and fixed the otherwise arbitrary normalization of the solutions. From the definition of the Jacobi zeta function, one gets the derivative of the solutions with respect to $\bar z$,
\begin{equation}
\psi'_\pm(\bar z)=[Z(\bar z\pm u,k)-Z(\bar z,k)\mp\Xi(u,k)]\psi_\pm(\bar z).
\end{equation}

We are now ready to apply the Gelfand-Yaglom theorem. This requires enclosing the system in a finite box, $-L\leq\bar z\leq+L$, and finding a solution $\varphi(\bar z)$ to the Lam\'e equation~\eqref{Hamaux2} satisfying the initial conditions $\varphi(-L)=0$, $\varphi'(-L)=1$. Such a solution is easily constructed from the expressions for $\psi_\pm(\bar z)$ and $\psi'_\pm(\bar z)$. What we need is then the value of this solution at $\bar z=+L$,
\begin{equation}
\varphi(L)=\frac1{\Delta(L,k)}\biggl[\frac{\Theta(L+u,k)}{\Theta(L-u,k)}e^{-2L\Xi(u,k)}-\frac{\Theta(L-u,k)}{\Theta(L+u,k)}e^{2L\Xi(u,k)}\biggr],
\label{phiL}
\end{equation}
where
\begin{equation}
\Delta(L,k)\equiv Z(L+u,k)-Z(L-u,k)-2\Xi(u,k).
\end{equation}
This looks rather complicated, but we should keep in mind that we are going to take the logarithm of the functional determinant, and are really only interested in its leading part for large $L$,
\begin{equation}
\log\varphi(L)=2L\Xi(u,k)+\mathcal O(L^0).
\end{equation}

We still need to evaluate in the same manner the analogous functional determinant for the normal phase, which corresponds to taking $\omega_n^2+\vek p_\perp^2+m_\pi^2-(m_\pi^2/k^2)\de_{\bar z}^2$ instead of eq.~\eqref{Hamaux}. This is equivalent to replacing eq.~\eqref{Hamaux2} with the eigenvalue condition $-\de_{\bar z}^2\psi=-\Omega_0^2\psi$. In this case, the linearly independent solutions are simply $e^{\pm\Omega_0\bar z}$. The equivalent of eq.~\eqref{phiL} for the free particle is therefore
\begin{equation}
\varphi_0(L)=\frac1{2\Omega_0}\left(e^{2L\Omega_0}-e^{-2L\Omega_0}\right).
\end{equation}
The Gelfand-Yaglom theorem then tells us that the logarithm of the regularized fluctuation determinant is simply
\begin{equation}
\log\frac{\varphi(L)}{\varphi_0(L)}=2L[\Xi(u,k)-\Omega_0]+\mathcal O(L^0).
\end{equation}
Upon returning to eq.~\eqref{F1CSL} and recalling that the size of the system in the rescaled variable $\bar z=zm_\pi/k$ is $2L$, we arrive at the conclusion that the one-loop free energy of neutral pion fluctuations above the CSL state, relative to the normal phase, equals
\begin{equation}\label{FGelfand}
\frac{\beta\Fa_\text{1,CSL}^{\text{1-loop},(\pi^0)}}V=\frac12\frac{m_\pi}k\left(\frac{e^{\gamma_\text{E}}\Lambda_\text{RG}^2}{4\pi}\right)^\epsilon\sum_n\int\frac{\dd^{d-1}\vek p_\perp}{(2\pi)^{d-1}}[\Xi(u,k)-\Omega_0].
\end{equation}
This is eq.~\eqref{monster} where the difference of the integrals in the curly brackets has been replaced with the expression~\eqref{Gelfand}.

We now have two different derivations of the one-loop free energy of neutral pions, leading to eqs.~\eqref{monster} and~\eqref{FGelfand}. The equivalence of these two expressions can be shown directly using the residue theorem. In particular, the parametric expression for the dispersion relation of the $n=1$ Lam\'e Hamiltonian~\eqref{SpecLamePar} can be used to convert the integrals in the curly brackets of~\eqref{monster} into an integral over a closed curve in the complex plane. It turns out that the integrand has poles in the interior of this integration curve and the residue gives exactly the expression~\eqref{Gelfand}.


\section{Next-to-leading-order results in the chiral limit}\label{app:chiral}

The limit of vanishing pion mass often leads to considerable simplification of computations. Also, it is a good first approximation when the pion mass is much smaller than other physical scales of the system such as the temperature or magnetic field. With this in mind, we work out here the NLO free energy of the CSL in the chiral limit.

This can in principle be extracted from our results valid for nonzero $m_\pi$ by taking simultaneously the limit $m_\pi \to 0$ and $k\to 0$ while keeping the ratio $m_\pi/k$ fixed according to eq.~\eqref{condkLargeB}. This procedure is, however, not completely straightforward. It appears easier to repeat the calculation of the free energy from scratch, assuming the chiral limit from the outset. Here one can benefit from the fact that in the chiral limit, the spectra of neutral pion fluctuations above the QCD vacuum and the CSL state are identical. Thus, the entire difference of the NLO free energies of the two states comes from the charged pion sector.

Specifically, the zero-temperature part of the renormalized NLO free energy becomes
\begin{align}\notag
\frac{\Fa_{\text{1,CSL}}^{T=0}}{V}\biggr|_{m_\pi= 0} = & \frac{(\phi_0')^4}{32\pi^2} \left(\log \frac{2B}{\Lambda_\text{RG}^2} - \frac{\bar{l}_1}{3} - \frac{2\,\bar{l}_2}{3}\right)\\
& + \frac{B^2}{4\pi^2}\left[\zeta^{(1,0)}\bigl(-1,\tfrac{1}{2}-\tfrac{(\phi_0')^2}{2B}\bigr)-\zeta^{(1,0)}\bigl(-1,\tfrac{1}{2}\bigr)\right],
\label{FT0chir}
\end{align}
where the superscripts of $\zeta$ refer to the number of derivatives of the Hurwitz $\zeta$-function with respect to its first and second argument. Also,
\begin{equation}
\phi_0'=\frac{\mu B}{4\pi^2 f_\pi^2}
\end{equation}
is the gradient of the CSL ground state~\eqref{GroundChir}. Note that we can strictly speaking no longer use the values of the finite counterterms $\bar l_i$, fixed in section~\ref{sec:renormalization}, since those were obtained at the renormalization scale $\Lambda_{\text{RG}}=m_\pi$. In the chiral limit, it appears sensible to choose $\Lambda_{\text{RG}}=f_\pi$, which sets the characteristic scale of $\chi$PT and is of the same order of magnitude as the physical value of $m_\pi$. In order to be able to explore the properties of the NLO free energy in the chiral limit numerically, we nevertheless resorted to the values of $\bar l_i$ shown in eq.~\eqref{Lbar} as a first estimate of the counterterms in the chiral limit. We found that in different regions of the parameter space, the sign of eq.~\eqref{FT0chir} varies, yet its sum with the LO free energy~\eqref{ECSLchir} remains negative. This confirms the finding made at LO that in the chiral limit, the CSL state is favored over the QCD vacuum for any nonzero magnetic field and baryon chemical potential. As an aside, the contribution~\eqref{FT0chir} picks an imaginary part if the magnetic field increases above the critical value~\eqref{BBECchiral}, but is finite at $B=B_\text{BEC}$.

Finally, the thermal part of the NLO free energy in the chiral limit reads
\begin{align}
\frac{\Fa_{\text{1,CSL}}^{T}}{V}\biggr|_{m_\pi= 0} = & \frac{B T}{2\pi^2} \sum_{m=0}^\infty \int \dd p_z \Bigl\{\log\left[1-e^{-\beta \epsilon_{\text{CSL}}(m,p_z)}\right] - \log\left[1-e^{-\beta \epsilon_{0}(m,p_z)}\right] \Bigr\},
\label{FTchir}
\end{align}
where this time
\begin{equation}
\begin{split}
\epsilon_{\text{CSL}}(m,p_z)&\equiv \sqrt{(2m+1)B+ p_z^2 - (\phi_0')^2},\\
\epsilon_0(m,p_z)&\equiv \sqrt{(2m+1)B+p_z^2}.
\end{split}
\label{defEpsBchir}
\end{equation}
The contribution of eq.~\eqref{FTchir} is manifestly negative for any $\mu$, $T$ and $B\leq B_\text{BEC}$ and diverges in the limit $B\to B_\text{BEC}$.


\bibliographystyle{JHEP}
\bibliography{references}

\providecommand{\href}[2]{#2}\begingroup\raggedright\begin{thebibliography}{10}

\bibitem{Brauner2017a}
T.~Brauner and N.~Yamamoto, \emph{{Chiral Soliton Lattice and Charged Pion
  Condensation in Strong Magnetic Fields}},
  \href{https://doi.org/10.1007/JHEP04(2017)132}{\emph{JHEP} {\bfseries 04}
  (2017) 132} [\href{https://arxiv.org/abs/1609.05213}{{\ttfamily
  1609.05213}}].

\bibitem{Kishine2015a}
J.-i.~Kishine and A.S.~Ovchinnikov, \emph{Theory of monoaxial chiral
  helimagnet}, \href{https://doi.org/10.1016/bs.ssp.2015.05.001}{\emph{Solid
  State Physics} {\bfseries 66} (2015) 1}.

\bibitem{Son2008a}
D.T.~Son and M.A.~Stephanov, \emph{{Axial anomaly and magnetism of nuclear and
  quark matter}}, \href{https://doi.org/10.1103/PhysRevD.77.014021}{\emph{Phys.
  Rev. D} {\bfseries 77} (2008) 014021}
  [\href{https://arxiv.org/abs/0710.1084}{{\ttfamily 0710.1084}}].

\bibitem{Eto:2012qd}
M.~Eto, K.~Hashimoto and T.~Hatsuda, \emph{{Ferromagnetic neutron stars: axial
  anomaly, dense neutron matter, and pionic wall}},
  \href{https://doi.org/10.1103/PhysRevD.88.081701}{\emph{Phys. Rev. D}
  {\bfseries 88} (2013) 081701}
  [\href{https://arxiv.org/abs/1209.4814}{{\ttfamily 1209.4814}}].

\bibitem{Yoshiike:2015tha}
R.~Yoshiike, K.~Nishiyama and T.~Tatsumi, \emph{{Spontaneous magnetization of
  quark matter in the inhomogeneous chiral phase}},
  \href{https://doi.org/10.1016/j.physletb.2015.10.028}{\emph{Phys. Lett. B}
  {\bfseries 751} (2015) 123}
  [\href{https://arxiv.org/abs/1507.02110}{{\ttfamily 1507.02110}}].

\bibitem{Eto:2022lhu}
M.~Eto and M.~Nitta, \emph{{Quantum nucleation of topological solitons}},
  \href{https://doi.org/10.1007/JHEP09(2022)077}{\emph{JHEP} {\bfseries 09}
  (2022) 077} [\href{https://arxiv.org/abs/2207.00211}{{\ttfamily
  2207.00211}}].

\bibitem{Higaki:2022gnw}
T.~Higaki, K.~Kamada and K.~Nishimura, \emph{{Formation of a chiral soliton
  lattice}}, \href{https://doi.org/10.1103/PhysRevD.106.096022}{\emph{Phys.
  Rev. D} {\bfseries 106} (2022) 096022}
  [\href{https://arxiv.org/abs/2207.00212}{{\ttfamily 2207.00212}}].

\bibitem{Lee2015a}
T.-G.~Lee, E.~Nakano, Y.~Tsue, T.~Tatsumi and B.~Friman, \emph{{Landau-Peierls
  instability in a Fulde-Ferrell type inhomogeneous chiral condensed phase}},
  \href{https://doi.org/10.1103/PhysRevD.92.034024}{\emph{Phys. Rev.}
  {\bfseries D92} (2015) 034024}
  [\href{https://arxiv.org/abs/1504.03185}{{\ttfamily 1504.03185}}].

\bibitem{Hidaka2015b}
Y.~Hidaka, K.~Kamikado, T.~Kanazawa and T.~Noumi, \emph{{Phonons, pions and
  quasi-long-range order in spatially modulated chiral condensates}},
  \href{https://doi.org/10.1103/PhysRevD.92.034003}{\emph{Phys. Rev.}
  {\bfseries D92} (2015) 034003}
  [\href{https://arxiv.org/abs/1505.00848}{{\ttfamily 1505.00848}}].

\bibitem{Ferrer2020}
E.J.~Ferrer and V.~de~la Incera, \emph{{Absence of Landau-Peierls Instability
  in the Magnetic Dual Chiral Density Wave Phase of Dense QCD}},
  \href{https://doi.org/10.1103/PhysRevD.102.014010}{\emph{Phys. Rev. D}
  {\bfseries 102} (2020) 014010}
  [\href{https://arxiv.org/abs/1902.06810}{{\ttfamily 1902.06810}}].

\bibitem{Brauner:2021sci}
T.~Brauner, H.~Kole\v{s}ov\'a and N.~Yamamoto, \emph{{Chiral soliton lattice
  phase in warm QCD}},
  \href{https://doi.org/10.1016/j.physletb.2021.136767}{\emph{Phys. Lett. B}
  {\bfseries 823} (2021) 136767}
  [\href{https://arxiv.org/abs/2108.10044}{{\ttfamily 2108.10044}}].

\bibitem{Wess1971a}
J.~Wess and B.~Zumino, \emph{{Consequences of anomalous Ward identities}},
  \href{https://doi.org/10.1016/0370-2693(71)90582-X}{\emph{Phys. Lett. B}
  {\bfseries 37} (1971) 95}.

\bibitem{Witten1983a}
E.~Witten, \emph{{Global aspects of current algebra}},
  \href{https://doi.org/10.1016/0550-3213(83)90063-9}{\emph{Nucl. Phys. B}
  {\bfseries 223} (1983) 422}.

\bibitem{Goldstone1981a}
J.~Goldstone and F.~Wilczek, \emph{{Fractional Quantum Numbers on Solitons}},
  \href{https://doi.org/10.1103/PhysRevLett.47.986}{\emph{Phys. Rev. Lett.}
  {\bfseries 47} (1981) 986}.

\bibitem{Scherer2012a}
S.~Scherer and M.R.~Schindler, \emph{{A Primer for Chiral Perturbation
  Theory}}, vol.~830 of \emph{Lecture Notes in Physics}, Springer (2012),
  \href{https://doi.org/10.1007/978-3-642-19254-8}{10.1007/978-3-642-19254-8}.

\bibitem{Andersen2012b}
J.O.~Andersen, \emph{{Chiral perturbation theory in a magnetic background -
  finite-temperature effects}},
  \href{https://doi.org/10.1007/JHEP10(2012)005}{\emph{JHEP} {\bfseries 10}
  (2012) 005} [\href{https://arxiv.org/abs/1205.6978}{{\ttfamily 1205.6978}}].

\bibitem{Gasser1984a}
J.~Gasser and H.~Leutwyler, \emph{Chiral perturbation theory to one loop},
  \href{https://doi.org/10.1016/0003-4916(84)90242-2}{\emph{Ann. Phys.}
  {\bfseries 158} (1984) 142}.

\bibitem{Ananthanarayan:2000ht}
B.~Ananthanarayan, G.~Colangelo, J.~Gasser and H.~Leutwyler, \emph{{Roy
  equation analysis of pi pi scattering}},
  \href{https://doi.org/10.1016/S0370-1573(01)00009-6}{\emph{Phys. Rept.}
  {\bfseries 353} (2001) 207}
  [\href{https://arxiv.org/abs/hep-ph/0005297}{{\ttfamily hep-ph/0005297}}].

\bibitem{Colangelo:2001df}
G.~Colangelo, J.~Gasser and H.~Leutwyler, \emph{{$\pi \pi$ scattering}},
  \href{https://doi.org/10.1016/S0550-3213(01)00147-X}{\emph{Nucl. Phys. B}
  {\bfseries 603} (2001) 125}
  [\href{https://arxiv.org/abs/hep-ph/0103088}{{\ttfamily hep-ph/0103088}}].

\bibitem{Baron:2011sf}
{\scshape ETM} collaboration, \emph{{Light hadrons from Nf=2+1+1 dynamical
  twisted mass fermions}},
  \href{https://doi.org/10.22323/1.105.0123}{\emph{PoS} {\bfseries LATTICE2010}
  (2010) 123} [\href{https://arxiv.org/abs/1101.0518}{{\ttfamily 1101.0518}}].

\bibitem{Aoki:2019cca}
{\scshape Flavour Lattice Averaging Group} collaboration, \emph{{FLAG Review
  2019: Flavour Lattice Averaging Group (FLAG)}},
  \href{https://doi.org/10.1140/epjc/s10052-019-7354-7}{\emph{Eur. Phys. J. C}
  {\bfseries 80} (2020) 113}
  [\href{https://arxiv.org/abs/1902.08191}{{\ttfamily 1902.08191}}].

\bibitem{Whittaker1927a}
E.T.~Whittaker and G.N.~Watson, \emph{A Course of Modern Analysis}, Cambridge
  University Press, Cambridge, UK (1927).

\bibitem{Agasian2001a}
N.O.~Agasian and I.A.~Shushpanov, \emph{{Gell-Mann-Oakes-Renner relation in a
  magnetic field at finite temperature}},
  \href{https://doi.org/10.1088/1126-6708/2001/10/006}{\emph{JHEP} {\bfseries
  10} (2001) 006} [\href{https://arxiv.org/abs/hep-ph/0107128}{{\ttfamily
  hep-ph/0107128}}].

\bibitem{Son2004a}
D.T.~Son and A.R.~Zhitnitsky, \emph{{Quantum anomalies in dense matter}},
  \href{https://doi.org/10.1103/PhysRevD.70.074018}{\emph{Phys. Rev. D}
  {\bfseries 70} (2004) 074018}
  [\href{https://arxiv.org/abs/hep-ph/0405216}{{\ttfamily hep-ph/0405216}}].

\bibitem{Li2000a}
H.~Li, D.~Kusnezov and F.~Iachello, \emph{{Group theoretical properties and
  band structure of the Lam\'e Hamiltonian}}, {\emph{J. Phys. A: Math. Gen.}
  {\bfseries 33} (2000) 6413}.

\bibitem{Dunne2008a}
G.V.~Dunne, \emph{{Functional determinants in quantum field theory}},
  \href{https://doi.org/10.1088/1751-8113/41/30/304006}{\emph{J. Phys. A}
  {\bfseries 41} (2008) 304006}
  [\href{https://arxiv.org/abs/0711.1178}{{\ttfamily 0711.1178}}].

\bibitem{Maier2008a}
R.S.~Maier, \emph{Lam{\'e} polynomials, hyperelliptic reductions and {Lam{\'e}}
  band structure}, \href{https://doi.org/10.1098/rsta.2007.2063}{\emph{Philos.
  Trans. Roy. Soc. A} {\bfseries 366} (2008) 1115}.

\bibitem{Brauner2017b}
T.~Brauner and S.~Kadam, \emph{{Anomalous electrodynamics of neutral pion
  matter in strong magnetic fields}},
  \href{https://doi.org/10.1007/JHEP03(2017)015}{\emph{JHEP} {\bfseries 03}
  (2017) 015} [\href{https://arxiv.org/abs/1701.06793}{{\ttfamily
  1701.06793}}].

\bibitem{Evans:2022hwr}
G.W.~Evans and A.~Schmitt, \emph{{Chiral anomaly induces superconducting baryon
  crystal}}, \href{https://doi.org/10.1007/JHEP09(2022)192}{\emph{JHEP}
  {\bfseries 09} (2022) 192}
  [\href{https://arxiv.org/abs/2206.01227}{{\ttfamily 2206.01227}}].

\bibitem{Abramowitz1972a}
M.~Abramowitz and I.A.~Stegun, \emph{Handbook of Mathematical Functions with
  Formulas, Graphs, and Mathematical Tables}, National Bureau of Standards
  (1972).

\bibitem{Tatsumi:2014wka}
T.~Tatsumi, K.~Nishiyama and S.~Karasawa, \emph{{Novel Lifshitz point for
  chiral transition in the magnetic field}},
  \href{https://doi.org/10.1016/j.physletb.2015.02.033}{\emph{Phys. Lett. B}
  {\bfseries 743} (2015) 66} [\href{https://arxiv.org/abs/1405.2155}{{\ttfamily
  1405.2155}}].

\bibitem{Abuki:2018iqp}
H.~Abuki, \emph{{Chiral crystallization in an external magnetic background:
  Chiral spiral versus real kink crystal}},
  \href{https://doi.org/10.1103/PhysRevD.98.054006}{\emph{Phys. Rev. D}
  {\bfseries 98} (2018) 054006}
  [\href{https://arxiv.org/abs/1808.05767}{{\ttfamily 1808.05767}}].

\bibitem{Nishiyama:2015fba}
K.~Nishiyama, S.~Karasawa and T.~Tatsumi, \emph{{Hybrid chiral condensate in
  the external magnetic field}},
  \href{https://doi.org/10.1103/PhysRevD.92.036008}{\emph{Phys. Rev. D}
  {\bfseries 92} (2015) 036008}
  [\href{https://arxiv.org/abs/1505.01928}{{\ttfamily 1505.01928}}].

\bibitem{Ferrer2015a}
E.J.~Ferrer and V.~de~la Incera, \emph{{Novel Topological Effects in Dense QCD
  in a Magnetic Field}},
  \href{https://doi.org/10.1016/j.nuclphysb.2018.04.009}{\emph{Nucl. Phys. B}
  {\bfseries 931} (2018) 192}
  [\href{https://arxiv.org/abs/1512.03972}{{\ttfamily 1512.03972}}].

\bibitem{Huang2018a}
X.-G.~Huang, K.~Nishimura and N.~Yamamoto, \emph{{Anomalous effects of dense
  matter under rotation}},
  \href{https://doi.org/10.1007/JHEP02(2018)069}{\emph{JHEP} {\bfseries 02}
  (2018) 069} [\href{https://arxiv.org/abs/1711.02190}{{\ttfamily
  1711.02190}}].

\bibitem{Nishimura:2020odq}
K.~Nishimura and N.~Yamamoto, \emph{{Topological term, QCD anomaly, and the
  $\eta^{'}$ chiral soliton lattice in rotating baryonic matter}},
  \href{https://doi.org/10.1007/JHEP07(2020)196}{\emph{JHEP} {\bfseries 07}
  (2020) 196} [\href{https://arxiv.org/abs/2003.13945}{{\ttfamily
  2003.13945}}].

\bibitem{Eto:2021gyy}
M.~Eto, K.~Nishimura and M.~Nitta, \emph{{Phases of rotating baryonic matter:
  non-Abelian chiral soliton lattices, antiferro-isospin chains, and
  ferri/ferromagnetic magnetization}},
  \href{https://doi.org/10.1007/JHEP08(2022)305}{\emph{JHEP} {\bfseries 08}
  (2022) 305} [\href{https://arxiv.org/abs/2112.01381}{{\ttfamily
  2112.01381}}].

\bibitem{Gronli:2022cri}
M.S.~Gr\o{}nli and T.~Brauner, \emph{{Competition of chiral soliton lattice and
  Abrikosov vortex lattice in QCD with isospin chemical potential}},
  \href{https://doi.org/10.1140/epjc/s10052-022-10300-5}{\emph{Eur. Phys. J. C}
  {\bfseries 82} (2022) 354}
  [\href{https://arxiv.org/abs/2201.07065}{{\ttfamily 2201.07065}}].

\bibitem{Adhikari:2022cks}
P.~Adhikari, E.~Leeser and J.~Markowski, \emph{{Phonon modes of magnetic vortex
  lattices in finite isospin QCD}},
  \href{https://arxiv.org/abs/2205.13369}{{\ttfamily 2205.13369}}.

\bibitem{Brauner2019c}
T.~Brauner, G.~Filios and H.~Kole\v{s}ov\'{a}, \emph{{Anomaly-Induced
  Inhomogeneous Phase in Quark Matter without the Sign Problem}},
  \href{https://doi.org/10.1103/PhysRevLett.123.012001}{\emph{Phys. Rev. Lett.}
  {\bfseries 123} (2019) 012001}
  [\href{https://arxiv.org/abs/1902.07522}{{\ttfamily 1902.07522}}].

\bibitem{Brauner2019b}
T.~Brauner, G.~Filios and H.~Kole\v{s}ov\'a, \emph{{Chiral soliton lattice in
  QCD-like theories}},
  \href{https://doi.org/10.1007/JHEP12(2019)029}{\emph{JHEP} {\bfseries 12}
  (2019) 029} [\href{https://arxiv.org/abs/1905.11409}{{\ttfamily
  1905.11409}}].

\end{thebibliography}\endgroup

\end{document}